\documentclass[%
 reprint,
 amsmath,amssymb,
 aps,
]{revtex4-2}

\usepackage{graphicx}% Include figure files
\usepackage{dcolumn}% Align table columns on decimal point
\usepackage{bm}% bold math
\usepackage{xcolor}
\begin{document}

\preprint{APS/123-QED}

\title{Probing valence electron and hydrogen dynamics using charge-pair imaging with ultrafast electron diffraction}% Force line breaks with \\
%with
%\affiliation{%

\author{Tianyu Wang\textsuperscript{1,2,*}}
\author{Hui Jiang\textsuperscript{1,2,3,*,\dag}}
\author{Ming Zhang\textsuperscript{4,*,\dag}}
\author{Xiao Zou\textsuperscript{1,2}}
\author{Pengfei Zhu\textsuperscript{1,3}}
\author{Feng He\textsuperscript{1}}
\author{Zheng Li\textsuperscript{4,5,6\dag}}
\author{Dao Xiang\textsuperscript{1,2,3,\dag}}

\affiliation{\textsuperscript{1}Key Laboratory for Laser Plasmas (Ministry of Education) and School of Physics and Astronomy, Collaborative Innovation Center for IFSA (CICIFSA), Shanghai Jiao Tong University, Shanghai 200240, China}
\affiliation{\textsuperscript{2}Zhangjiang Institute for Advanced Study, Shanghai Jiao Tong University, Shanghai 201210, China}
\affiliation{\textsuperscript{3}Tsung-Dao Lee Institute, Shanghai Jiao Tong University, Shanghai 201210, China}
\affiliation{\textsuperscript{4}State Key Laboratory for Mesoscopic Physics and Center for Nano-Optoelectronics, School of Physics, Peking University, Beijing 100871, China}
\affiliation{\textsuperscript{5}Collaborative Innovation Center of Extreme Optics, Shanxi University, Taiyuan, Shanxi 030006, China}

\affiliation{\textsuperscript{6}Peking University Yangtze Delta Institute of Optoelectronics, Nantong, Jiangsu 226010, China}

\thanks{These authors contributed equally to this work.}
\thanks{\textsuperscript{\dag}Corresponding authors: jianghui1997@sjtu.edu.cn, zhming@alumni.pku.edu.cn, zheng.li@pku.edu.cn, dxiang@sjtu.edu.cn}

%              %  but any date may be explicitly specified

\begin{abstract}
%This study presents time-resolved real-space tracking of valence electron and proton dynamics during the photodissociation of ammonia using MeV ultrafast electron diffraction with a temporal resolution of 130 fs FWHM. Crucially, we adapt the charge pair distribution function, originally used for liquid-phase structure retrieval, to disentangle the valence electron dynamics and electron correlation from the inelastic scattering, and to extract real-space electron density alongside nuclear positions. We resolve the correlated motion of valence electrons and protons in the adiabatic and non-adiabatic dissociation channels of ammonia, demonstrating a direct visualization and unprecedented details of coupled electronic and nuclear dynamics in photoexcited ammonia molecule.
A key challenge in ultrafast science has been to directly track the coupled motions of electrons and nuclei in real-space and real-time. This study presents a significant step towards this goal by demonstrating the feasibility of time-resolved real-space tracking of valence electron and hydrogen dynamics during the photodissociation of ammonia (NH$_3$) using MeV ultrafast electron diffraction. It is demonstrated that the enhanced temporal resolution, in conjunction with the analysis of the charge-pair distribution function, enables the disentanglement of the correlated motion of valence electrons and hydrogens in photoexcited ammonia molecule. The methodology employed in this study, which utilizes the charge-pair distribution function from ultrafast electron scattering to retrieve intertwined electron and nucleus dynamics, may open up new opportunities in the study of quantum dynamics for a wide range of molecules. 
\end{abstract}

\maketitle

%\section{\label{sec:level1} Introduction}

The redistribution of valence electrons upon photoexcitation, coupled with subsequent atomic motions, determines the outcome of photophysical and photochemical processes~\cite{zewail2000}. Thus, tracking electronic and nuclear dynamics simultaneously in real-time and real-space is of the utmost importance to gain a deep understanding of the fundamental mechanisms behind these photoinduced phenomena~\cite{goulielmakis2010,hockett2011}. The realization of this goal became increasingly attainable with the advent of ultrafast lasers, which offer high temporal resolution, and bright x-ray and electron beams, which provide high spatial resolution. The high brightness and ultrashort pulse durations of x-ray beams in free-electron lasers (FEL)~\cite{Emma4:NP10, FELreview, LCLSreview, RPP} and electron beams in MeV ultrafast electron diffraction (UED)~\cite{Weathersby86:RSI15, Qi2020,Millerreview, UEDreview} have enabled the study of many prototypical photochemical reactions with hard x-ray FEL~\cite{CHD266LCLS, I2LCLS, stankus2019, ruddock_deep_2019, yong_observation_2020, yong_ultrafast_2021} and MeV UED~\cite{YangJ16:PRL117, YangJ18:Scicen361, Wolf19:NatChem504, YangJ20:Science368,wang_imaging_2025, green_imaging_2025} facilities. While it is broadly acknowledged that ultrafast scattering with x-rays and electrons primarily provides information of molecular structure, recent studies have demonstrated the sensitivity to excited electronic dynamics. In ultrafast x-ray scattering, where x-rays are exclusively scattered from the electrons of a molecule, a clear signature of electron density redistribution has been observed~\cite{yong_observation_2020}. This observation has been made immediately following photoexcitation, prior to the onset of substantial structural change that could otherwise obscure the signal. In ultrafast electron scattering, where the incoming electrons are scattered by both the electrons and nuclei of the molecules, an inelastic scattering signal has been observed~\cite{YangJ20:Science368, ThomasWolf131:PRL23, wang_imaging_2025, green_imaging_2025}. The inelastic signal reflects the change of the electron correlation upon pump excitation from $S_0$ to $S_1$ state, which originates from two-electron effects including Coulomb repulsion and Pauli exclusion~\cite{YangJ20:Science368}. However, a comprehensive analysis of the evolution of the spatial distribution of both the valence electrons and nuclei in a photochemical reaction has yet to be accomplished.  

\begin{figure}
    \centering
    \includegraphics[width=0.8\linewidth]{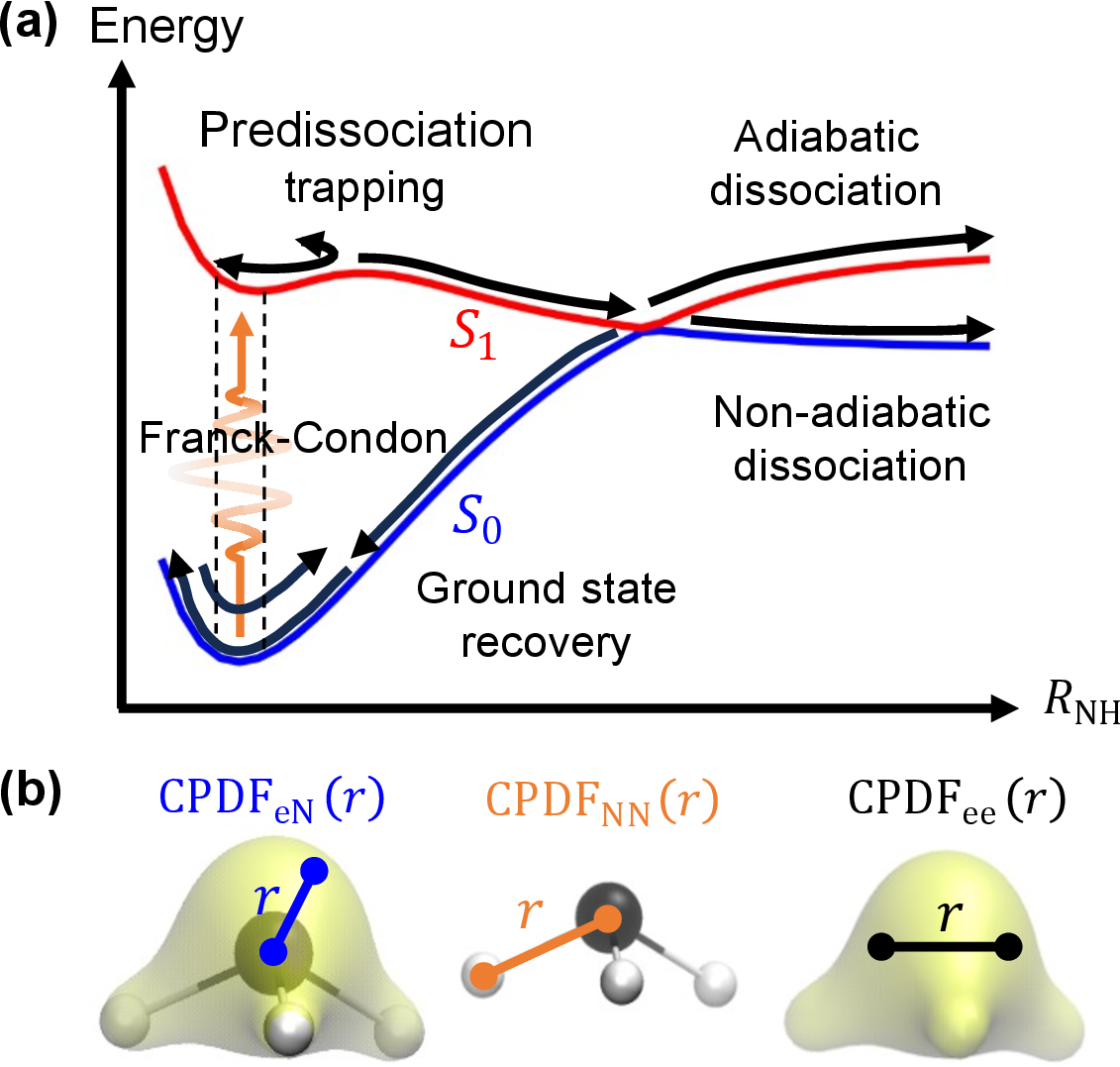}
    \caption{(a) Schematic of photochemical dynamics in NH$_3$. (b) Charge-pair distribution functions (CPDF), contributed by  electron-nucleus  CPDF$_{\mathrm{eN}}(r)$, nucleus-nucleus CPDF$_{\mathrm{NN}}(r)$ and electron-electron CPDF$_{\mathrm{ee}}(r)$, where $r$ is the distance between the charge-pair. Black and white spheres represent C and H atoms, respectively, and the isosurface correspond to the electron density of NH$_3$ equilibrium geometry in $S_0$ state. }
    %(a) Schematic of NH$_3$ valence electron and proton dynamics triggered by photoexcitation. The potential energy curves (PEC) of ground state $S_0$ and excited state $S_1$ are plotted along the elongation of one N--H bond in planar geometry, with four reaction channels predissociation trapping, adiabatic dissociation, non-adiabatic dissociation and ground-state recovery. (b) Charge-pair distribution functions (CPDF), and three types of charge-pairs: electrons and nuclei CPDF$_{\mathrm{eN}}(r)$, nuclei and nuclei CPDF$_{\mathrm{NN}}(r)$, electrons and electrons CPDF$_{\mathrm{ee}}(r)$, where $r$ is the distance between the specific charge-pair. Black and white spheres represent C and H atoms, respectively, and the isosurface correspond to the electron density of NH$_3$ equilibrium geometry in $S_0$ state.}
    \label{fig:schematic}
\end{figure}

In this Letter, we demonstrate a critical step towards this goal by demonstrating the capability to track both the electronic and nuclear dynamics in photodissocation of ammonia (NH$_3$)~\cite{ThomasWolf131:PRL23,Zhu137:JCP12,Ma137:JCP12,Zotev16:JCTC20,Rodríguez16:PCCP14,Xie142:JCP15,Hause12:JCP06,McCarthy86:JCP87,Li114:JPCA10}. Despite the fact that the number of core electrons is markedly small, rendering ammonia an ideal platform for the study of valence electron dynamics, the small scattering cross section and the fast motion of the lightest hydrogen atom result in significant challenges in the experiment. In a previous study with MeV UED~\cite{ThomasWolf131:PRL23}, the temporal resolution (approximately 500 fs FWHM) was compromised to achieve sufficient scattering signal by increasing the charge of the electron pulse. This, in turn, led to broadening of the electron pulse due to the action of the Coulomb repulsion force. Furthermore, deuterated ammonia (ND$_3$) with slower dynamics was utilized in the experiment, and the photodissociation dynamics were not tracked in real-space due to the contamination of the inelastic signal, which complicates the determination of the pair distribution function (PDF). In this study, we harness the negative dispersion of a double-bend achromatic (DBA) lens to compress the elongated electron beam, enabling the generation of a high-charge electron beam with short pulse width~\cite{Qi2020}. This results in a temporal resolution of approximately 130 fs while still maintaining a moderate scattering signal, allowing the observation of photodissociation dynamics for NH$_3$ without deuteration in real-time. Moreover, drawing inspiration from the charge-pair distribution function (CPDF), which was originally developed for structure retrieval in liquid phase scattering~\cite{YangJ23:PCCP21}, we were able to disentangle the valence electron and hydrogen dynamics in real-space, circumventing the difficulty imposed by the inelastic scattering signal, which partially overlaps with the elastic scattering signal.

In our experiment, the NH$_3$ molecules are photoexcited from ground state $S_0$ to the first excited state $S_1$ by a 201.1 nm pump laser with a pulse width of approximately 120 fs. Given that the equilibrium geometry of NH$_3$ is pyramidal in $S_0$ state, but planar in $S_1$ state, the umbrella vibration mode ($\nu_2'=4$) is strongly excited at this wavelength~\cite{Cheng647:Astro06}. As illustrated in Fig.~1(a), after leaving the Franck-Condon region, the excited ammonia molecules may proceed in the $S_1$ state with four reaction channels.
Most NH$_3$ molecules return to the ground state $S_0$ by non-adiabatic coupling via conical intersections, bounded with high vibrational excitations or dissociate to NH$_2$+H.
Other molecules remain in the excited state $S_1$, either dissociates adiabatically or trapped by the potential barrier of the $S_1$ state.
%(1) predissociation trapping, i.e., the NH$_3$ molecules are temporarily confined behind the potential barrier before dissociation in $S_1$; (2) adiabatic dissociation along the excited state $S_1$; (3) non-adiabatic dissociation via the conical intersection; (4) ground state recovery, namely, molecules decay to the ground state with high vibrational excitation. The dissociation channels (2) and (3) yield NH$_2$+H fragments. 
%Apart from the N--H bond elongation, the excited-state dynamics involve large-amplitude umbrella motion,
%Distinct from the equilibrium pyramidal geometry in $S_0$ state, the planar geometry in $S_1$ state has lower energy, thus the umbrella motion is strongly excited. 
%The channels of the photodissociation dynamics of NH$_3$ is shown in . 
%The photochemical dynamics of ammonia in the $S_1$ state has four reaction channels (see Fig.~\ref{fig:schematic}(a)): (1) predissociation trapping, i.e., the NH$_3$ molecules are temporarily confined behind the potential barrier before dissociation in $S_1$; (2) adiabatic dissociation along the excited state $S_1$; (3) non-adiabatic dissociation via the conical intersection; (4) ground state recovery, namely, molecules decay to the ground state with high vibrational excitation. The dissociation channels (2) and (3) yield NH$_2$+H fragments.

%

Conventionally, the study of structural dynamics has been undertaken through the utilization of PDF~\cite{YangJ16:PRL117,YangJ18:Scicen361,YangJ20:Science368,YangJ21:Nature596,Wolf19:NatChem504,  wang_imaging_2025, green_imaging_2025, PDF}, which denotes the probability of finding an atomic pair at a specified distance. However, PDF is based on independent atom model (IAM), which requires the separation of atomic and molecular scattering signals~\cite{YangJ23:PCCP21, PDF}. 
%IAM neglects the redistribution of electrons due to bond formation and does not take into account the electron correlation effects that may lead to considerable inelastic scattering signals~\cite{YangJ20:Science368}. Consequently, the PDF analysis method is suited for molecules with heavy atoms, where the dominant scattering signals originate from core electrons and are largely unaltered by bond formation. However, for NH$_3$ molecule where the electron density is significantly affected by bond formation, the results from PDF analysis may not be reliable. In this study, we utilize the CPDF approach to account for the impact of electron redistribution. 
Beyond IAM model, CPDF (see End Matter for definition) takes into account the redistribution of electrons due to bond formation and electron correlation effects that may lead to considerable inelastic scattering signals. Therefore, IAM model and PDF analysis is suited for molecules with heavy atoms, where the dominant scattering signals originate from core electrons and are largely unaltered by bond formation. However, in NH$_3$, the electron density is significantly affected by bond formation, so we utilize the CPDF approach to account for the impact of electron redistribution. 
As shown in Fig.~\ref{fig:schematic}(b), CPDF describes the probability distribution of each charge-pair in the molecule, including electron-nucleus CPDF$_{\mathrm{eN}}(r)$, nucleus-nucleus CPDF$_{\mathrm{NN}}(r)$ and electron-electron CPDF$_{\mathrm{ee}}(r)$. We demonstrate that the combination of the three components can sensitively reflect the valence electron dynamics. 
The experiment was performed with the MeV-UED instrument at Shanghai Jiao Tong University~\cite{Qi2020}. The electron beam charge was increased from the nominal value of 20 fC to approximately 40 fC to compensate for the low scattering cross section of NH$_3$. The DBA compressor facilitates the generation of an ultrashort electron pulse (approximately 50 fs FWHM) with negligible timing jitter, thereby ensuring an overall instrument response function (IRF) of approximately 130 fs (FWHM) in this measurement. The molecules are delivered by a pulsed nozzle operating at 400 Hz. At the interaction point, the UV laser spot size is approximately 250~$\mu$m, and the pulse energy is approximately 12~$\mu$J to ensure a low excitation ratio (approximately 1.5\%). Each scattering pattern at a designated time delay is accumulated over a period of 10 seconds with 4000 electron pulses. The diffraction pattern is measured with a time step of 33~fs in the time window from $-0.5$~ps to 0.5~ps and the time step is 100 fs from 0.5 ps to 1.0 ps. The full data sets include 4700 scans, with the total integration time at each time delay amounting to approximately 13 hours. %A phosphor screen is employed to capture the diffraction pattern, which is then imaged onto an electron-multiplying charge-coupled device (EMCCD) camera. The phosphor screen contains a 3-millimeter diameter hole in the center, and consequently, the data within the range of $s<0.8~\text{\AA}^{-1}$ are not measured. Each scattering pattern at a designated time delay is accumulated over a period of 10 seconds with 4000 electron pulses. The diffraction pattern is measured with a time step of 33~fs in the time window from $-0.5$~ps to 0.5~ps and the time step is 100 fs from 0.5 ps to 1.0 ps. The full data sets include 3600 scans, with the total integration time at each time delay amounting to approximately 10 hours. % and the total data acquisition time amounting to approximately 16 days. 

\begin{figure*}
    \centering
    \includegraphics[width=0.9\linewidth]{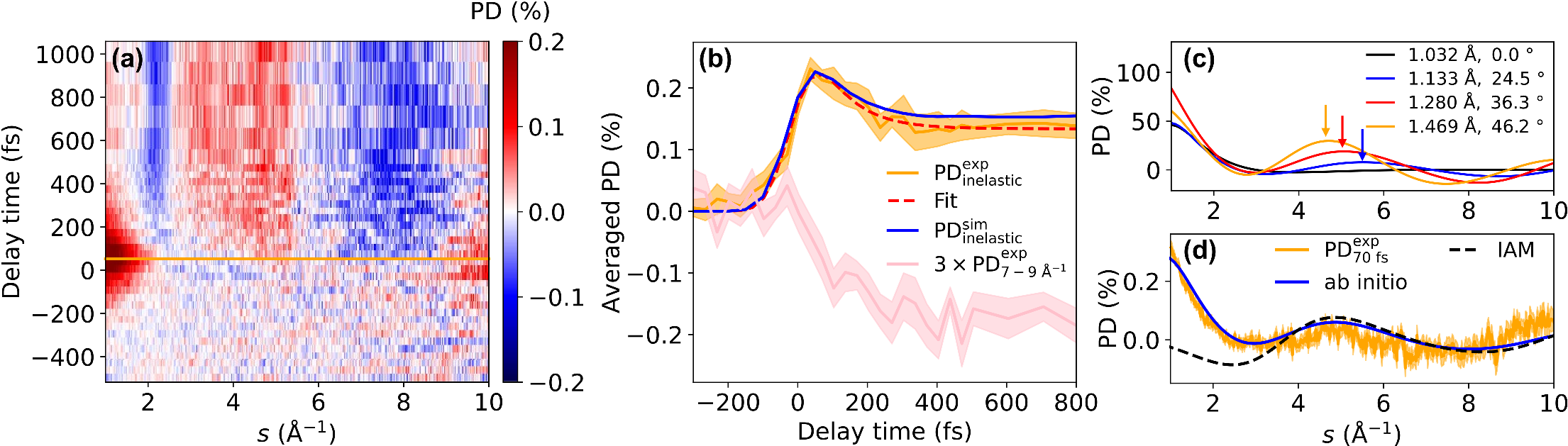}
    \caption{
    Momentum-space analysis of the diffraction signal. (a) Time-resolved experimental percent-difference (PD) signal. The orange line marks the PD at the time when the inelastic scattering signal reaches its maximum (70~fs). (b) Temporal evolution of averaged PD over selected $s$-ranges. The orange and pink curves correspond to inelastic (1–1.5~$\text{\AA}^{-1}$) and elastic (7–9~$\text{\AA}^{-1}$) contributions, respectively. The red dashed line is a fit with exponential decay convolved with the instrument response function. The blue line shows the total inelastic signal from \textit{ab initio} simulation. (c) \textit{Ab initio} calculated PD signals for various NH$_3$ geometries associated with the umbrella mode. The corresponding N–H bond lengths and pyramid angles are labeled. (d) Comparison of the PD at 70~fs (orange), marked in (a), with simulated PDs from \textit{ab initio} (blue) and IAM (black) calculations, averaged over geometries sampled along the umbrella mode. Shaded region indicates 68\% confidence interval.
    %The analysis of diffraction signal in momentum space. (a) Time-resolved measurement of percentage difference (PD) signal.
    %(b) Temporal evolution of the averaged PD signal over different $s$-ranges. The orange curve shows the experimental averaged PD signals corresponding to inelastic contributions within 1--1.5~$\text{\AA}^{-1}$, obtained by subtracting the extrapolated elastic component from the total signal. The red dashed line represents a fitted curve obtained by convolving an exponential decay plus a constant background with the experimental time resolution. The blue line shows the total inelastic signal from the \textit{ab initio} simulation. The pink curve shows the experimental averaged PD signals corresponding to elastic contributions within 7--9~$\text{\AA}^{-1}$. Theoretical results are convolved with the experimental temporal resolution.
    %(c) \textit{Ab initio} calculated PD for a series of geometries along the umbrella mode in the $S_1$ state. For each geometry, the N--H bond length and pyramid angle are labeled. 
    %(d) The orange line represents the transient PD feature observed at 70~fs in the experimental measurement. The blue and black lines represent PD signals calculated from \textit{ab initio} and IAM models, respectively, averaged over geometries sampled along the umbrella mode in panel (c). The shaded area around the experimental data represents the 68\% confidence interval estimated by bootstrap resampling.
    }
    \label{fig:PD_exp}
\end{figure*}

The measured time-resolved diffraction signals are given as percent-difference (PD), as shown in Fig.~2(a). The PD signal is characterized by a narrow enhanced band at $s<2~\text{\AA}^{-1}$, a narrow bleached band in the region $2<s<2.5~\text{\AA}^{-1}$, a wide enhanced band at $2.5<s<5.5~\text{\AA}^{-1}$,  and a wide bleached band at $s>5.5~\text{\AA}^{-1}$. In order to understand the experimental PD signal, we perform molecular dynamics (MD) simulation of NH$_3$ molecules using the fewest-switches surface hopping (FSSH) method, based on the potential energy surface in Ref.~\cite{Zhu137:JCP12}. From the MD trajectories, we calculate the UED signal both \textit{ab initio} and within IAM model (see End Matter and Fig.~\ref{fig:PD_IAM_ab} for details of MD simulations and calculation of UED signals). Similar to the simulation results reported in Ref.~\cite{ThomasWolf131:PRL23}, the IAM, which neglects both the inelastic scattering signal stemming from electron-electron correlation and the elastic scattering signal resulting from electron redistribution, failed to account for the measured enhanced band at $s<2~\text{\AA}^{-1}$. In contrast, simulation with \textit{ab initio} electron scattering reproduced the enhanced band at $s<2~\text{\AA}^{-1}$ which is primarily due to inelastic scattering. %PD is defined as
%\begin{equation}
%    {\rm{PD}}(s,t)=\frac{I(s,t)-I(s,t<0)}{I(s,t<0)}\,,
%        \label{cal_PD}
%\end{equation}
%where $I(s,t)$ is the measured scattering signal at momentum transfer $s$ and time delay $t$, and  $I(s,t<0)$ is the signal when the electron beam arrives at the sample earlier than the pump laser. The PD signal is characterized by a narrow enhanced band at $s<2~\text{\AA}^{-1}$, a narrow bleached band in the region $2<s<2.5~\text{\AA}^{-1}$, a wide enhanced band at $2.5<s<5.5~\text{\AA}^{-1}$,  and a wide bleached band at $s>5.5~\text{\AA}^{-1}$. 

The measured PD signal is in qualitative agreement with that reported in Ref.~\cite{ThomasWolf131:PRL23}. However, the higher temporal resolution of the present measurement allows for the retrieval of more detailed information regarding the dynamics. As shown in Fig.~\ref{fig:PD_exp}(b), the measured inelastic scattering signal (see End Matter for details regarding the subtraction of the elastic scattering signal) can be adequately fitted with a convolution of the IRF (130~fs) and an exponential decay function with a constant background, yielding a decay constant of
96$\pm22$ fs. This constant is presumably associated with the lifetime of the excited electronic state and is comparable to values reported in previous studies~\cite{Wells130:JCP09}. The temporal profile of measured inelastic scattering signal is in good agreement with the \textit{ab initio} calculation results, as shown in Fig.~\ref{fig:PD_exp}(b). Moreover, the enhanced temporal resolution also facilitates the observation of the earlier onset of the inelastic scattering signal in the low $s$ region where the electronic state information is encoded, as opposed to that in the high $s$ region (pink line in Fig.~\ref{fig:PD_exp}(b)), which is predominantly associated with structural changes.

Due to the short lifetime of the excited state, it is challenging to capture the predissociation dynamics for ammonia~\cite{ThomasWolf131:PRL23}. 
Given the proportional relationship between the inelastic signal and the population of the predissociation channel~\cite{YangJ20:Science368}, we anticipate that the PD signal at around 70~fs (indicated by the orange line in Fig.~\ref{fig:PD_exp}(a)) when the inelastic signal is maximal may contain the signature for predissociation. As shown in Fig.~2(d), in addition to the pronounced positive signal at low $s$ region, the PD signal at 70~fs also exhibits a weak positive peak near 5~$\text{\AA}^{-1}$. 
This peak may be qualitatively associated with an approximate bond length of ${2\pi}/{5~\mathrm{\AA}^{-1}}\approx1.26~\text{\AA}$. 
To understand the observed feature, in Fig.~\ref{fig:PD_exp}(c), we present the \textit{ab initio} calculated PD signals in $S_1$ state (corresponding to 100\% excitation ratio) for different NH$_3$ geometries associated with the umbrella vibrational mode~\cite{Zotev16:JCTC20}. We find that the positions of these peaks (as indicated by the arrows in Fig.~\ref{fig:PD_exp}(c)) closely match the values estimated by $2\pi/R_\text{N-H}$, and the simulated PD can reproduce the measured peak near 5~$\text{\AA}^{-1}$ when the N-H bond length is approximately 1.28~$\text{\AA}$. In order to account for the time resolution, the simulated PD signals have been averaged over different geometries, as shown in Fig.~\ref{fig:PD_exp}(d). The excellent agreement between the \textit{ab initio} simulation and the experimental measurement indicates that the umbrella-mode transient structure in the predissociation stage has been captured. In contrast, the simulated PD signals from IAM reproduce the measured results solely in the $s>4~\text{\AA}^{-1}$ range, where the signal is predominantly influenced by structural changes. The substantial discrepancy between the measurement and IAM calculation underscores the considerable influence of valence electron dynamics in the measurement.

\begin{figure*}
    \centering
    \includegraphics[width=0.9\linewidth]{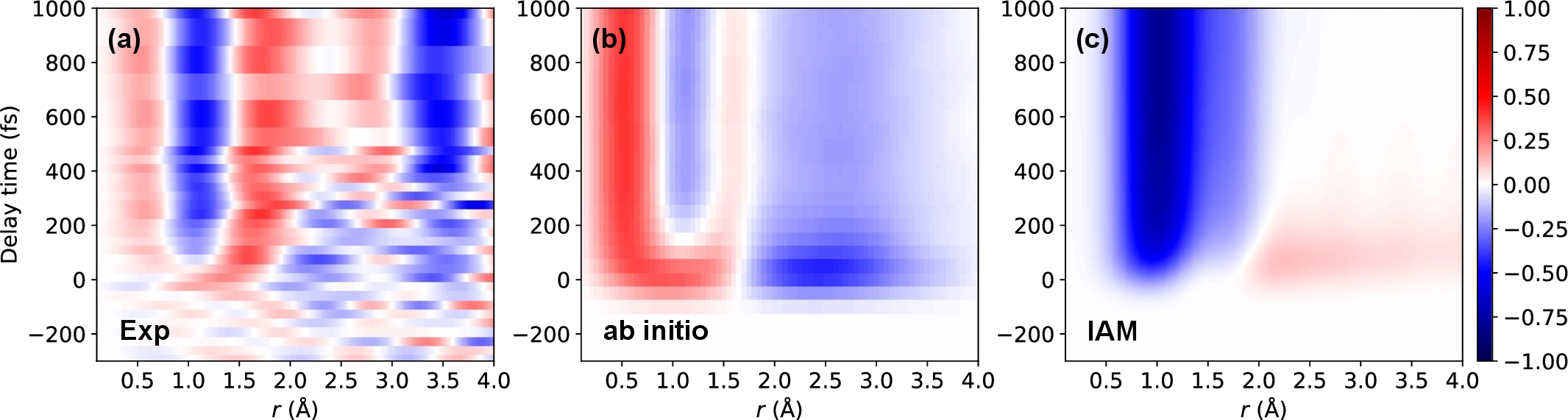}
    \caption{Analysis of difference charge-pair distribution function ($\Delta$CPDF) in real-space and real-time. (a) Experimental $\Delta$CPDF. (b) Theoretical $\Delta$CPDF obtained from \textit{ab initio} diffraction intensity based on the molecular dynamics trajectories. (c) Theoretical $\Delta$PDF obtained with the IAM model. The theoretical results are convolved with a Gaussian function with 130-fs FWHM.
    }
    \label{fig:CPDF_exp}
\end{figure*}

A distinctive benefit of UED is that the electrons' wavelength is considerably short, enabling the measurement of the scattering signal over a broad momentum transfer range. This, in turn, facilitates precise inversion to retrieve the information in real-space. However, conventional PDF analysis based on IAM requires the removal of low-angle inelastic scattering signals, which introduces substantial truncation errors in the low-$s$ region and compromises information pertaining to valence electrons. The aforementioned issues can be circumvented by employing CPDF which inverts the full scattering signal to recover charge-pair information in real-space. Moreover, given that electrons and hydrogen nuclei carry the same unit charge, the changes in the scattering signal induced by valence electron motion are comparable to those from hydrogen atoms. Consequently, the CPDF facilitates the concurrent imaging of both valence electrons and hydrogen dynamics.

Fig.~\ref{fig:CPDF_exp}(a) shows the inverted experimental $\Delta$CPDF. It should be noted that the interpretation of the positive and negative values in the $\Delta$CPDF is dependent on the product of the charges. For electron–electron (ee) and nucleus–nucleus (NN) pairs, a positive signal indicate the increase in pair density at a certain distance. By contrast, for electron–nucleus (eN) pairs, due to their opposite charges, a positive $\Delta$CPDF value indicates a decrease in pair density. The most distinct features of the measured $\Delta$CPDF are a bleached band near 1~$\text{\AA}$ and “U”-shaped enhanced bands surrounding it. Given that the dominant signal in NH$_3$ dissociation originates from the breaking of the N–H bond at approximately 1~$\text{\AA}$, the negative band that emerges at this distance at around 100 fs signifies the decrease in pair density for N–H NN pairs. 

To gain a more profound understanding of the features observed in the experimental $\Delta$CPDF, the theoretical $\Delta$CPDF is calculated from \textit{ab initio} simulations (Fig.~\ref{fig:CPDF_exp}(b)) and the theoretical $\Delta$PDF is derived from IAM calculations (Fig.~\ref{fig:CPDF_exp}(c)). The \textit{ab initio} $\Delta$CPDF in Fig.~\ref{fig:CPDF_exp}(b) also exhibited a “U”-shaped positive feature near 1~$\text{\AA}$, in good agreement with the measurement. In contrast, the IAM-derived $\Delta$PDF in Fig.~\ref{fig:CPDF_exp}(c) exhibited solely a broad negative signal below 2~$\text{\AA}$ and a broad positive signal above 2~$\text{\AA}$, a typical feature for dissociation. As illustrated in Fig.~\ref{fig:CPDF_exp}(b), a negative band between 2 and 4~$\text{\AA}$ is also evident, peaking around time zero and decaying as the time delay increases. It is evident that the ground-state structure of NH$_3$ contains virtually no charge-pairs beyond 2~$\text{\AA}$. Therefore, any signal change in this region, whether positive or negative, must originate from the formation of new charge-pairs, and only eN pairs can contribute negative signals as pair density increases. Because of the nuclei's comparatively slower movement relative to the electrons, the pronounced negative signal around time zero likely arises from the motion of the electrons. %Furthermore, a close comparison between Fig.~\ref{fig:CPDF_exp}(a) and Fig.~\ref{fig:CPDF_exp}(b) indicate that the positive and negative signals in the 2.5--4~$\text{\AA}$ region observed in experiment were missing in the \textit{ab initio} calculation. 

\begin{figure*}
    \centering
    \includegraphics[width=0.9\linewidth]{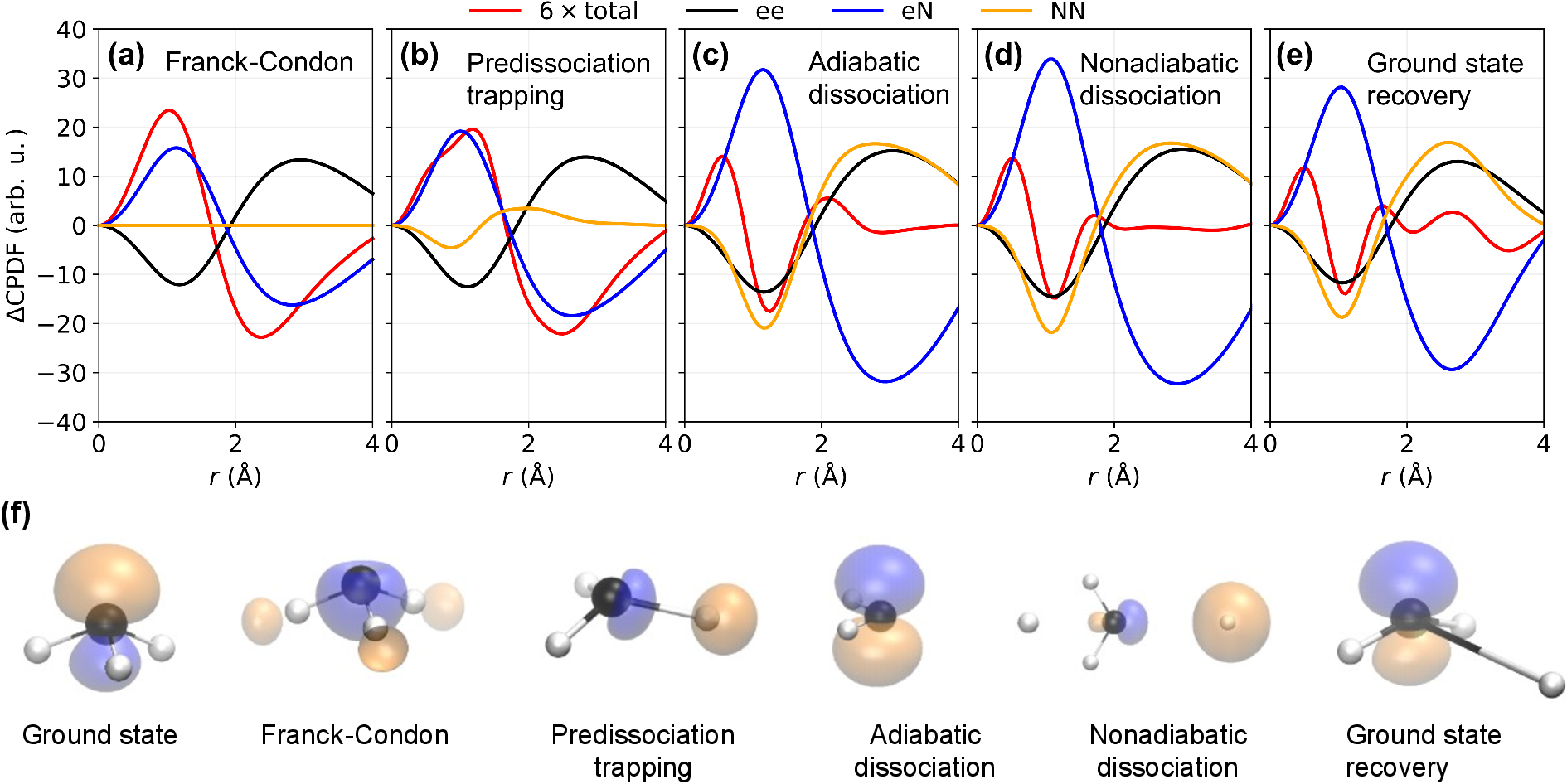}
    \caption{Decomposition of $\Delta$CPDF into different reaction channels. (a)--(e) The red, black, blue, and orange lines represent the total $\Delta$CPDF, $\Delta$CPDF$_{\text{ee}}$, $\Delta$CPDF$_{\text{eN}}$, and $\Delta$CPDF$_{\text{NN}}$, respectively. The total $\Delta$CPDF (red line) is scaled by a factor of 6. The corresponding reaction channels are labeled in each panel. (f) Schematic of representative molecular structures and the associated electronic orbitals, including the non-bonding orbital $n$ of N atom and the anti-bonding orbital $\sigma^*$ of N--H bond.
    }
    \label{fig:CPDF_sim}
\end{figure*}

It is worth mentioning that the $\Delta$CPDF includes contributions from all reaction channels and from all ee, eN, and NN pairs, thereby complicating the identification of the predominant contribution to each feature. Here we exploit the unique advantage of theory over experiment to decompose the signal into different parts. Figs.~\ref{fig:CPDF_sim}(a)-(e) show representative decompositions of the $\Delta$CPDF for different reaction channels with separated contribution from ee, eN, and NN pairs. The representative molecular structures and associated electronic orbitals are schematically shown in Fig.~\ref{fig:CPDF_sim}(f). 

We firstly discuss the signal around time zero. In the Franck-Condon region, the pump pulse excites an electron from the non-bonding orbital $n$ to the anti-bonding orbital $\sigma^*$, leading to more diffuse distribution for the electron cloud. Because the nuclei has not yet moved, the signal from NN pair is zero (orange line in Fig.~4(a)). The rapid redistribution of the electrons results in a decrease in density for the ee and eN pairs at about 1~$\text{\AA}$ (the original N–H bond length). Given that the contribution is dominated by eN pairs (blue line in Fig.~4(a)), the decrease in density, however, yields a positive signal at 1~$\text{\AA}$, as can be seen with the red line in Fig.~\ref{fig:CPDF_sim}(a). Meanwhile, the diffused electron cloud increases the density of eN pairs at large distances, resulting in a negative signal at $r>2~\text{\AA}$. All these features are in excellent agreement with the measured $\Delta$CPDF near time zero as shown in Fig.~\ref{fig:CPDF_exp}(a), providing strong support that the motion of the valence electrons, which occurs prior to substantial structural alterations, has been successfully captured.

After leaving the Franck-Condon region, the molecule quickly passes into the predissociation state, where the structural changes become more pronounced. The elongation of the N--H bonds results in a decrease of the NN density at 1~$\text{\AA}$. This, in turn, gives rise to a small negative signal at 1~$\text{\AA}$ (orange line in Fig.~4(b)), which compensates for the positive signal from eN pairs. As a result, the total intensity of the positive peak at 1~$\text{\AA}$ is reduced and the position of the peak shifts to larger distance in comparison to the Franck-Condon case. 
%and shift to the double-peak feature as shown in Fig.~\ref{fig:CPDF_sim}(c) and (d).

After crossing the dissociation barrier, the molecules proceed along the dissociation pathways. As illustrated in Figs.~\ref{fig:CPDF_sim}(c) and \ref{fig:CPDF_sim}(d), the breaking of the N-H bond results in the generation of a pronounced negative signal around 1~$\text{\AA}$, primarily contributed by the ee and NN pairs. In the region between 2.5 and 4 $\text{\AA}$, the positive contributions from ee and NN pairs are offset by the negative contribution from eN pairs, resulting in a weak negative signal (red lines in Figs.~4(c) and 4(d)). The differences between the two dissociation pathways are relatively minor, as the signal is dominated by the broken N-H bonds. However, due to changes in the electronic states, the valence electron orbitals differ, and the vibrational energies of the nuclei also vary. These differences may become measurable in the future with a high repetition rate MeV UED with enhanced signal-to-noise ratio. Nevertheless, our measurement is capable of resolving the full dynamics of photexcited ammonia, ranging from the redistribution of valence electrons to the breaking of N-H bonds. The characteristics of the simulated signals in the Franck–Condon region and predissociation state, together with that for the dissociation channels, provide a coherent explanation for the “U”-shaped feature observed in the experiment. 
%and can only be distinguished by taking their difference. This analysis is presented in the Appendix.

According to the MD simulation, only a small fraction of the molecules return to the ground state without dissociating (the ground state recovery channel). These molecules possess substantial kinetic energy, thereby inducing enhanced nuclear vibrations that increase the probability of finding atom pairs at large distances. As a result, this results in a modulation of the $\Delta$CPDF between 2 and 4~$\text{\AA}$ (red line in Fig.~4(e)), which closely matches those observed in the experiment. Notably, the positive and negative signals at 2.5-4~$\text{\AA}$ in Fig.~\ref{fig:CPDF_exp}(a) emerge only after about 200 fs following photoexcitation, which can be attributed to the delayed onset of the ground state recovery pathway. The discrepancy between the experimental observations (Fig.~3(a)) and the simulation results (Fig.~3(b)) in the region 2.5-4~$\text{\AA}$ suggests that the MD simulation may have underestimated the population of the ground state recovery channel.

In summary, we have demonstrated the feasibility of studying valence electron and hydrogen dynamics simultaneously in real-space and real-time using UED. Through analysis of CPDF, we have captured the full dynamics of valence electrons and hydrogens during the NH$_3$ dissociation reaction, including the motion of valence electrons relative to nuclei at the moment of photoexcitation, umbrella-like vibrations of the hydrogen atoms before dissociaton, N-H bond cleavage during dissociation, and vibration of the hot molecules in the ground state. It is important to note that the $\Delta$CPDF$_{\text{ee}}$ term is equivalent to the difference in radial distribution function as measured by X-ray scattering. As shown in Figs.~\ref{fig:CPDF_sim}a-e, the difference of $\Delta$CPDF$_{\text{ee}}$ is minor among the various channels. In contrast, additional contributions from $\Delta$CPDF$_{\text{eN}}$ and $\Delta$CPDF$_{\text{NN}}$ in electron scattering result in distinguishable features in the total $\Delta$CPDF for different reaction pathways, as the nuclei provide a natural coordinate reference for locating the valence electrons. We anticipate that the demonstrated methodology may offer new insights into the coupled electronic and nuclear dynamics for a wide range of molecules.

\begin{acknowledgments}
This work is supported by the National Natural Science Foundation of China (Grant No. 12525501, 12335010, 12174009, 12234002 and 92250303), the National Key R\&D Program of China (Grant No. 2023YFA1406801 and 2024YFA1612204) and the Beijing Natural Science Foundation (Grant No. Z220008). D.X. would like to acknowledge the support from the New Cornerstone Science Foundation through the Xplorer Prize. The UED experiment was supported by the Shanghai soft x-ray free-electron laser facility.
\end{acknowledgments}

%\nocite{*}

\bibliography{ref}% Produces the bibliography via BibTeX.

\begin{thebibliography}{39}%
\makeatletter
\providecommand \@ifxundefined [1]{%
 \@ifx{#1\undefined}
}%
\providecommand \@ifnum [1]{%
 \ifnum #1\expandafter \@firstoftwo
 \else \expandafter \@secondoftwo
 \fi
}%
\providecommand \@ifx [1]{%
 \ifx #1\expandafter \@firstoftwo
 \else \expandafter \@secondoftwo
 \fi
}%
\providecommand \natexlab [1]{#1}%
\providecommand \enquote  [1]{``#1''}%
\providecommand \bibnamefont  [1]{#1}%
\providecommand \bibfnamefont [1]{#1}%
\providecommand \citenamefont [1]{#1}%
\providecommand \href@noop [0]{\@secondoftwo}%
\providecommand \href [0]{\begingroup \@sanitize@url \@href}%
\providecommand \@href[1]{\@@startlink{#1}\@@href}%
\providecommand \@@href[1]{\endgroup#1\@@endlink}%
\providecommand \@sanitize@url [0]{\catcode `\\12\catcode `\$12\catcode `\&12\catcode `\#12\catcode `\^12\catcode `\_12\catcode `\%12\relax}%
\providecommand \@@startlink[1]{}%
\providecommand \@@endlink[0]{}%
\providecommand \url  [0]{\begingroup\@sanitize@url \@url }%
\providecommand \@url [1]{\endgroup\@href {#1}{\urlprefix }}%
\providecommand \urlprefix  [0]{URL }%
\providecommand \Eprint [0]{\href }%
\providecommand \doibase [0]{https://doi.org/}%
\providecommand \selectlanguage [0]{\@gobble}%
\providecommand \bibinfo  [0]{\@secondoftwo}%
\providecommand \bibfield  [0]{\@secondoftwo}%
\providecommand \translation [1]{[#1]}%
\providecommand \BibitemOpen [0]{}%
\providecommand \bibitemStop [0]{}%
\providecommand \bibitemNoStop [0]{.\EOS\space}%
\providecommand \EOS [0]{\spacefactor3000\relax}%
\providecommand \BibitemShut  [1]{\csname bibitem#1\endcsname}%
\let\auto@bib@innerbib\@empty
%</preamble>
\bibitem [{\citenamefont {Zewail}(2000)}]{zewail2000}%
  \BibitemOpen
  \bibfield  {author} {\bibinfo {author} {\bibfnamefont {A.~H.}\ \bibnamefont {Zewail}},\ }\bibfield  {title} {\bibinfo {title} {Femtochemistry: {Atomic}-{Scale} {Dynamics} of the {Chemical} {Bond} {Using} {Ultrafast} {Lasers} ({Nobel} {Lecture})},\ }\href {https://doi.org/10.1002/1521-3773(20000804)39:15<2586::AID-ANIE2586>3.0.CO;2-O} {\bibfield  {journal} {\bibinfo  {journal} {Angewandte Chemie International Edition}\ }\textbf {\bibinfo {volume} {39}},\ \bibinfo {pages} {2586} (\bibinfo {year} {2000})}\BibitemShut {NoStop}%
\bibitem [{\citenamefont {Goulielmakis}\ \emph {et~al.}(2010)\citenamefont {Goulielmakis}, \citenamefont {Loh}, \citenamefont {Wirth}, \citenamefont {Santra}, \citenamefont {Rohringer}, \citenamefont {Yakovlev}, \citenamefont {Zherebtsov}, \citenamefont {Pfeifer}, \citenamefont {Azzeer}, \citenamefont {Kling}, \citenamefont {Leone},\ and\ \citenamefont {Krausz}}]{goulielmakis2010}%
  \BibitemOpen
  \bibfield  {author} {\bibinfo {author} {\bibfnamefont {E.}~\bibnamefont {Goulielmakis}}, \bibinfo {author} {\bibfnamefont {Z.-H.}\ \bibnamefont {Loh}}, \bibinfo {author} {\bibfnamefont {A.}~\bibnamefont {Wirth}}, \bibinfo {author} {\bibfnamefont {R.}~\bibnamefont {Santra}}, \bibinfo {author} {\bibfnamefont {N.}~\bibnamefont {Rohringer}}, \bibinfo {author} {\bibfnamefont {V.~S.}\ \bibnamefont {Yakovlev}}, \bibinfo {author} {\bibfnamefont {S.}~\bibnamefont {Zherebtsov}}, \bibinfo {author} {\bibfnamefont {T.}~\bibnamefont {Pfeifer}}, \bibinfo {author} {\bibfnamefont {A.~M.}\ \bibnamefont {Azzeer}}, \bibinfo {author} {\bibfnamefont {M.~F.}\ \bibnamefont {Kling}}, \bibinfo {author} {\bibfnamefont {S.~R.}\ \bibnamefont {Leone}},\ and\ \bibinfo {author} {\bibfnamefont {F.}~\bibnamefont {Krausz}},\ }\bibfield  {title} {\bibinfo {title} {Real-time observation of valence electron motion},\ }\href {https://doi.org/10.1038/nature09212} {\bibfield  {journal} {\bibinfo  {journal} {Nature}\ }\textbf {\bibinfo {volume}
  {466}},\ \bibinfo {pages} {739} (\bibinfo {year} {2010})}\BibitemShut {NoStop}%
\bibitem [{\citenamefont {Hockett}\ \emph {et~al.}(2011)\citenamefont {Hockett}, \citenamefont {Bisgaard}, \citenamefont {Clarkin},\ and\ \citenamefont {Stolow}}]{hockett2011}%
  \BibitemOpen
  \bibfield  {author} {\bibinfo {author} {\bibfnamefont {P.}~\bibnamefont {Hockett}}, \bibinfo {author} {\bibfnamefont {C.~Z.}\ \bibnamefont {Bisgaard}}, \bibinfo {author} {\bibfnamefont {O.~J.}\ \bibnamefont {Clarkin}},\ and\ \bibinfo {author} {\bibfnamefont {A.}~\bibnamefont {Stolow}},\ }\bibfield  {title} {\bibinfo {title} {Time-resolved imaging of purely valence-electron dynamics during a chemical reaction},\ }\href {https://doi.org/10.1038/nphys1980} {\bibfield  {journal} {\bibinfo  {journal} {Nature Physics}\ }\textbf {\bibinfo {volume} {7}},\ \bibinfo {pages} {612} (\bibinfo {year} {2011})}\BibitemShut {NoStop}%
\bibitem [{\citenamefont {Emma}\ \emph {et~al.}(2010)\citenamefont {Emma}, \citenamefont {Akre}, \citenamefont {Arthur}, \citenamefont {Bionta}, \citenamefont {Bostedt}, \citenamefont {Bozek}, \citenamefont {Brachmann}, \citenamefont {Bucksbaum}, \citenamefont {Coffee}, \citenamefont {Decker}, \citenamefont {Ding}, \citenamefont {Dowell}, \citenamefont {Edstrom}, \citenamefont {Fisher}, \citenamefont {Frisch}, \citenamefont {Gilevich}, \citenamefont {Hastings}, \citenamefont {Hays}, \citenamefont {Hering}, \citenamefont {Huang}, \citenamefont {Iverson}, \citenamefont {Loos}, \citenamefont {Messerschmidt}, \citenamefont {Miahnahri}, \citenamefont {Moeller}, \citenamefont {Nuhn}, \citenamefont {Pile}, \citenamefont {Ratner}, \citenamefont {Rzepiela}, \citenamefont {Schultz}, \citenamefont {Smith}, \citenamefont {Stefan}, \citenamefont {Tompkins}, \citenamefont {Turner}, \citenamefont {Welch}, \citenamefont {White}, \citenamefont {Wu}, \citenamefont {Yocky},\ and\ \citenamefont {Galayda}}]{Emma4:NP10}%
  \BibitemOpen
  \bibfield  {author} {\bibinfo {author} {\bibfnamefont {P.}~\bibnamefont {Emma}}, \bibinfo {author} {\bibfnamefont {R.}~\bibnamefont {Akre}}, \bibinfo {author} {\bibfnamefont {J.}~\bibnamefont {Arthur}}, \bibinfo {author} {\bibfnamefont {R.}~\bibnamefont {Bionta}}, \bibinfo {author} {\bibfnamefont {C.}~\bibnamefont {Bostedt}}, \bibinfo {author} {\bibfnamefont {J.}~\bibnamefont {Bozek}}, \bibinfo {author} {\bibfnamefont {A.}~\bibnamefont {Brachmann}}, \bibinfo {author} {\bibfnamefont {P.}~\bibnamefont {Bucksbaum}}, \bibinfo {author} {\bibfnamefont {R.}~\bibnamefont {Coffee}}, \bibinfo {author} {\bibfnamefont {F.-J.}\ \bibnamefont {Decker}}, \bibinfo {author} {\bibfnamefont {Y.}~\bibnamefont {Ding}}, \bibinfo {author} {\bibfnamefont {D.}~\bibnamefont {Dowell}}, \bibinfo {author} {\bibfnamefont {S.}~\bibnamefont {Edstrom}}, \bibinfo {author} {\bibfnamefont {A.}~\bibnamefont {Fisher}}, \bibinfo {author} {\bibfnamefont {J.}~\bibnamefont {Frisch}}, \bibinfo {author} {\bibfnamefont {S.}~\bibnamefont {Gilevich}},
  \bibinfo {author} {\bibfnamefont {J.}~\bibnamefont {Hastings}}, \bibinfo {author} {\bibfnamefont {G.}~\bibnamefont {Hays}}, \bibinfo {author} {\bibfnamefont {P.}~\bibnamefont {Hering}}, \bibinfo {author} {\bibfnamefont {Z.}~\bibnamefont {Huang}}, \bibinfo {author} {\bibfnamefont {R.}~\bibnamefont {Iverson}}, \bibinfo {author} {\bibfnamefont {H.}~\bibnamefont {Loos}}, \bibinfo {author} {\bibfnamefont {M.}~\bibnamefont {Messerschmidt}}, \bibinfo {author} {\bibfnamefont {A.}~\bibnamefont {Miahnahri}}, \bibinfo {author} {\bibfnamefont {S.}~\bibnamefont {Moeller}}, \bibinfo {author} {\bibfnamefont {H.-D.}\ \bibnamefont {Nuhn}}, \bibinfo {author} {\bibfnamefont {G.}~\bibnamefont {Pile}}, \bibinfo {author} {\bibfnamefont {D.}~\bibnamefont {Ratner}}, \bibinfo {author} {\bibfnamefont {J.}~\bibnamefont {Rzepiela}}, \bibinfo {author} {\bibfnamefont {D.}~\bibnamefont {Schultz}}, \bibinfo {author} {\bibfnamefont {T.}~\bibnamefont {Smith}}, \bibinfo {author} {\bibfnamefont {P.}~\bibnamefont {Stefan}}, \bibinfo {author}
  {\bibfnamefont {H.}~\bibnamefont {Tompkins}}, \bibinfo {author} {\bibfnamefont {J.}~\bibnamefont {Turner}}, \bibinfo {author} {\bibfnamefont {J.}~\bibnamefont {Welch}}, \bibinfo {author} {\bibfnamefont {W.}~\bibnamefont {White}}, \bibinfo {author} {\bibfnamefont {J.}~\bibnamefont {Wu}}, \bibinfo {author} {\bibfnamefont {G.}~\bibnamefont {Yocky}},\ and\ \bibinfo {author} {\bibfnamefont {J.}~\bibnamefont {Galayda}},\ }\bibfield  {title} {\bibinfo {title} {First lasing and operation of an {\aa{}ngstrom}-wavelength free-electron laser},\ }\href@noop {} {\bibfield  {journal} {\bibinfo  {journal} {Nature Photonics}\ }\textbf {\bibinfo {volume} {4}},\ \bibinfo {pages} {641} (\bibinfo {year} {2010})}\BibitemShut {NoStop}%
\bibitem [{\citenamefont {Pellegrini}\ \emph {et~al.}(2016)\citenamefont {Pellegrini}, \citenamefont {Marinelli},\ and\ \citenamefont {Reiche}}]{FELreview}%
  \BibitemOpen
  \bibfield  {author} {\bibinfo {author} {\bibfnamefont {C.}~\bibnamefont {Pellegrini}}, \bibinfo {author} {\bibfnamefont {A.}~\bibnamefont {Marinelli}},\ and\ \bibinfo {author} {\bibfnamefont {S.}~\bibnamefont {Reiche}},\ }\bibfield  {title} {\bibinfo {title} {The physics of x-ray free-electron lasers},\ }\href {https://doi.org/10.1103/RevModPhys.88.015006} {\bibfield  {journal} {\bibinfo  {journal} {Rev. Mod. Phys.}\ }\textbf {\bibinfo {volume} {88}},\ \bibinfo {pages} {015006} (\bibinfo {year} {2016})}\BibitemShut {NoStop}%
\bibitem [{\citenamefont {Bostedt}\ \emph {et~al.}(2016)\citenamefont {Bostedt}, \citenamefont {Boutet}, \citenamefont {Fritz}, \citenamefont {Huang}, \citenamefont {Lee}, \citenamefont {Lemke}, \citenamefont {Robert}, \citenamefont {Schlotter}, \citenamefont {Turner},\ and\ \citenamefont {Williams}}]{LCLSreview}%
  \BibitemOpen
  \bibfield  {author} {\bibinfo {author} {\bibfnamefont {C.}~\bibnamefont {Bostedt}}, \bibinfo {author} {\bibfnamefont {S.}~\bibnamefont {Boutet}}, \bibinfo {author} {\bibfnamefont {D.~M.}\ \bibnamefont {Fritz}}, \bibinfo {author} {\bibfnamefont {Z.}~\bibnamefont {Huang}}, \bibinfo {author} {\bibfnamefont {H.~J.}\ \bibnamefont {Lee}}, \bibinfo {author} {\bibfnamefont {H.~T.}\ \bibnamefont {Lemke}}, \bibinfo {author} {\bibfnamefont {A.}~\bibnamefont {Robert}}, \bibinfo {author} {\bibfnamefont {W.~F.}\ \bibnamefont {Schlotter}}, \bibinfo {author} {\bibfnamefont {J.~J.}\ \bibnamefont {Turner}},\ and\ \bibinfo {author} {\bibfnamefont {G.~J.}\ \bibnamefont {Williams}},\ }\bibfield  {title} {\bibinfo {title} {Linac coherent light source: The first five years},\ }\href {https://doi.org/10.1103/RevModPhys.88.015007} {\bibfield  {journal} {\bibinfo  {journal} {Rev. Mod. Phys.}\ }\textbf {\bibinfo {volume} {88}},\ \bibinfo {pages} {015007} (\bibinfo {year} {2016})}\BibitemShut {NoStop}%
\bibitem [{\citenamefont {Seddon}\ \emph {et~al.}(2017)\citenamefont {Seddon}, \citenamefont {Clarke}, \citenamefont {Dunning}, \citenamefont {Masciovecchio}, \citenamefont {Milne}, \citenamefont {Parmigiani}, \citenamefont {Rugg}, \citenamefont {Spence}, \citenamefont {Thompson}, \citenamefont {Ueda}, \citenamefont {Vinko}, \citenamefont {Wark},\ and\ \citenamefont {Wurth}}]{RPP}%
  \BibitemOpen
  \bibfield  {author} {\bibinfo {author} {\bibfnamefont {E.~A.}\ \bibnamefont {Seddon}}, \bibinfo {author} {\bibfnamefont {J.~A.}\ \bibnamefont {Clarke}}, \bibinfo {author} {\bibfnamefont {D.~J.}\ \bibnamefont {Dunning}}, \bibinfo {author} {\bibfnamefont {C.}~\bibnamefont {Masciovecchio}}, \bibinfo {author} {\bibfnamefont {C.~J.}\ \bibnamefont {Milne}}, \bibinfo {author} {\bibfnamefont {F.}~\bibnamefont {Parmigiani}}, \bibinfo {author} {\bibfnamefont {D.}~\bibnamefont {Rugg}}, \bibinfo {author} {\bibfnamefont {J.~C.~H.}\ \bibnamefont {Spence}}, \bibinfo {author} {\bibfnamefont {N.~R.}\ \bibnamefont {Thompson}}, \bibinfo {author} {\bibfnamefont {K.}~\bibnamefont {Ueda}}, \bibinfo {author} {\bibfnamefont {S.~M.}\ \bibnamefont {Vinko}}, \bibinfo {author} {\bibfnamefont {J.~S.}\ \bibnamefont {Wark}},\ and\ \bibinfo {author} {\bibfnamefont {W.}~\bibnamefont {Wurth}},\ }\bibfield  {title} {\bibinfo {title} {Short-wavelength free-electron laser sources and science: a review},\ }\href
  {https://doi.org/10.1088/1361-6633/aa7cca} {\bibfield  {journal} {\bibinfo  {journal} {Rep. Prog. Phys.}\ }\textbf {\bibinfo {volume} {80}},\ \bibinfo {pages} {115901} (\bibinfo {year} {2017})}\BibitemShut {NoStop}%
\bibitem [{\citenamefont {Weathersby}\ \emph {et~al.}(2015)\citenamefont {Weathersby}, \citenamefont {Brown}, \citenamefont {Centurion}, \citenamefont {Chase}, \citenamefont {Coffee}, \citenamefont {Corbett}, \citenamefont {Eichner}, \citenamefont {Frisch}, \citenamefont {Fry}, \citenamefont {Guehr}, \citenamefont {Hartmann}, \citenamefont {Hast}, \citenamefont {Hettel}, \citenamefont {Jobe}, \citenamefont {Jongewaard}, \citenamefont {Lewandowski}, \citenamefont {Li}, \citenamefont {Lindenberg}, \citenamefont {Makasyuk}, \citenamefont {May}, \citenamefont {McCormick}, \citenamefont {Nguyen}, \citenamefont {Reid}, \citenamefont {Shen}, \citenamefont {Sokolowski-Tinten}, \citenamefont {Vecchione}, \citenamefont {Vetter}, \citenamefont {Wu}, \citenamefont {Yang}, \citenamefont {Duerr},\ and\ \citenamefont {Wang}}]{Weathersby86:RSI15}%
  \BibitemOpen
  \bibfield  {author} {\bibinfo {author} {\bibfnamefont {S.~P.}\ \bibnamefont {Weathersby}}, \bibinfo {author} {\bibfnamefont {G.}~\bibnamefont {Brown}}, \bibinfo {author} {\bibfnamefont {M.}~\bibnamefont {Centurion}}, \bibinfo {author} {\bibfnamefont {T.~F.}\ \bibnamefont {Chase}}, \bibinfo {author} {\bibfnamefont {R.}~\bibnamefont {Coffee}}, \bibinfo {author} {\bibfnamefont {J.}~\bibnamefont {Corbett}}, \bibinfo {author} {\bibfnamefont {J.~P.}\ \bibnamefont {Eichner}}, \bibinfo {author} {\bibfnamefont {J.~C.}\ \bibnamefont {Frisch}}, \bibinfo {author} {\bibfnamefont {A.~R.}\ \bibnamefont {Fry}}, \bibinfo {author} {\bibfnamefont {M.}~\bibnamefont {Guehr}}, \bibinfo {author} {\bibfnamefont {N.}~\bibnamefont {Hartmann}}, \bibinfo {author} {\bibfnamefont {C.}~\bibnamefont {Hast}}, \bibinfo {author} {\bibfnamefont {R.}~\bibnamefont {Hettel}}, \bibinfo {author} {\bibfnamefont {R.~K.}\ \bibnamefont {Jobe}}, \bibinfo {author} {\bibfnamefont {E.~N.}\ \bibnamefont {Jongewaard}}, \bibinfo {author} {\bibfnamefont
  {J.~R.}\ \bibnamefont {Lewandowski}}, \bibinfo {author} {\bibfnamefont {R.~K.}\ \bibnamefont {Li}}, \bibinfo {author} {\bibfnamefont {A.~M.}\ \bibnamefont {Lindenberg}}, \bibinfo {author} {\bibfnamefont {I.}~\bibnamefont {Makasyuk}}, \bibinfo {author} {\bibfnamefont {J.~E.}\ \bibnamefont {May}}, \bibinfo {author} {\bibfnamefont {D.}~\bibnamefont {McCormick}}, \bibinfo {author} {\bibfnamefont {M.~N.}\ \bibnamefont {Nguyen}}, \bibinfo {author} {\bibfnamefont {A.~H.}\ \bibnamefont {Reid}}, \bibinfo {author} {\bibfnamefont {X.}~\bibnamefont {Shen}}, \bibinfo {author} {\bibfnamefont {K.}~\bibnamefont {Sokolowski-Tinten}}, \bibinfo {author} {\bibfnamefont {T.}~\bibnamefont {Vecchione}}, \bibinfo {author} {\bibfnamefont {S.~L.}\ \bibnamefont {Vetter}}, \bibinfo {author} {\bibfnamefont {J.}~\bibnamefont {Wu}}, \bibinfo {author} {\bibfnamefont {J.}~\bibnamefont {Yang}}, \bibinfo {author} {\bibfnamefont {H.~A.}\ \bibnamefont {Duerr}},\ and\ \bibinfo {author} {\bibfnamefont {X.~J.}\ \bibnamefont {Wang}},\ }\bibfield
  {title} {\bibinfo {title} {Mega-electron-volt ultrafast electron diffraction at {SLAC} national accelerator laboratory},\ }\href@noop {} {\bibfield  {journal} {\bibinfo  {journal} {Rev. Sci. Instrum.}\ }\textbf {\bibinfo {volume} {86}} (\bibinfo {year} {2015})}\BibitemShut {NoStop}%
\bibitem [{\citenamefont {Qi}\ \emph {et~al.}(2020)\citenamefont {Qi}, \citenamefont {Ma}, \citenamefont {Zhao}, \citenamefont {Cheng}, \citenamefont {Jiang}, \citenamefont {Lu}, \citenamefont {Jiang}, \citenamefont {Qian}, \citenamefont {Wang}, \citenamefont {Zhang}, \citenamefont {Zhu}, \citenamefont {Zou}, \citenamefont {Wan}, \citenamefont {Xiang},\ and\ \citenamefont {Zhang}}]{Qi2020}%
  \BibitemOpen
  \bibfield  {author} {\bibinfo {author} {\bibfnamefont {F.}~\bibnamefont {Qi}}, \bibinfo {author} {\bibfnamefont {Z.}~\bibnamefont {Ma}}, \bibinfo {author} {\bibfnamefont {L.}~\bibnamefont {Zhao}}, \bibinfo {author} {\bibfnamefont {Y.}~\bibnamefont {Cheng}}, \bibinfo {author} {\bibfnamefont {W.}~\bibnamefont {Jiang}}, \bibinfo {author} {\bibfnamefont {C.}~\bibnamefont {Lu}}, \bibinfo {author} {\bibfnamefont {T.}~\bibnamefont {Jiang}}, \bibinfo {author} {\bibfnamefont {D.}~\bibnamefont {Qian}}, \bibinfo {author} {\bibfnamefont {Z.}~\bibnamefont {Wang}}, \bibinfo {author} {\bibfnamefont {W.}~\bibnamefont {Zhang}}, \bibinfo {author} {\bibfnamefont {P.}~\bibnamefont {Zhu}}, \bibinfo {author} {\bibfnamefont {X.}~\bibnamefont {Zou}}, \bibinfo {author} {\bibfnamefont {W.}~\bibnamefont {Wan}}, \bibinfo {author} {\bibfnamefont {D.}~\bibnamefont {Xiang}},\ and\ \bibinfo {author} {\bibfnamefont {J.}~\bibnamefont {Zhang}},\ }\bibfield  {title} {\bibinfo {title} {Breaking 50 femtosecond resolution barrier in mev
  ultrafast electron diffraction with a double bend achromat compressor},\ }\href {https://doi.org/10.1103/PhysRevLett.124.134803} {\bibfield  {journal} {\bibinfo  {journal} {Phys. Rev. Lett.}\ }\textbf {\bibinfo {volume} {124}},\ \bibinfo {pages} {134803} (\bibinfo {year} {2020})}\BibitemShut {NoStop}%
\bibitem [{\citenamefont {Ischenko}\ \emph {et~al.}(2017)\citenamefont {Ischenko}, \citenamefont {Weber},\ and\ \citenamefont {Miller}}]{Millerreview}%
  \BibitemOpen
  \bibfield  {author} {\bibinfo {author} {\bibfnamefont {A.~A.}\ \bibnamefont {Ischenko}}, \bibinfo {author} {\bibfnamefont {P.~M.}\ \bibnamefont {Weber}},\ and\ \bibinfo {author} {\bibfnamefont {R.~J.~D.}\ \bibnamefont {Miller}},\ }\bibfield  {title} {\bibinfo {title} {Capturing chemistry in action with electrons: Realization of atomically resolved reaction dynamics},\ }\href@noop {} {\bibfield  {journal} {\bibinfo  {journal} {Chem. Rev.}\ }\textbf {\bibinfo {volume} {117}},\ \bibinfo {pages} {11066} (\bibinfo {year} {2017})}\BibitemShut {NoStop}%
\bibitem [{\citenamefont {Filippetto}\ \emph {et~al.}(2022)\citenamefont {Filippetto}, \citenamefont {Musumeci}, \citenamefont {Li}, \citenamefont {Siwick}, \citenamefont {Otto}, \citenamefont {Centurion},\ and\ \citenamefont {Nunes}}]{UEDreview}%
  \BibitemOpen
  \bibfield  {author} {\bibinfo {author} {\bibfnamefont {D.}~\bibnamefont {Filippetto}}, \bibinfo {author} {\bibfnamefont {P.}~\bibnamefont {Musumeci}}, \bibinfo {author} {\bibfnamefont {R.~K.}\ \bibnamefont {Li}}, \bibinfo {author} {\bibfnamefont {B.~J.}\ \bibnamefont {Siwick}}, \bibinfo {author} {\bibfnamefont {M.~R.}\ \bibnamefont {Otto}}, \bibinfo {author} {\bibfnamefont {M.}~\bibnamefont {Centurion}},\ and\ \bibinfo {author} {\bibfnamefont {J.~P.~F.}\ \bibnamefont {Nunes}},\ }\bibfield  {title} {\bibinfo {title} {Ultrafast electron diffraction: Visualizing dynamic states of matter},\ }\href {https://doi.org/10.1103/RevModPhys.94.045004} {\bibfield  {journal} {\bibinfo  {journal} {Rev. Mod. Phys.}\ }\textbf {\bibinfo {volume} {94}},\ \bibinfo {pages} {045004} (\bibinfo {year} {2022})}\BibitemShut {NoStop}%
\bibitem [{\citenamefont {Minitti}\ \emph {et~al.}(2015)\citenamefont {Minitti}, \citenamefont {Budarz}, \citenamefont {Kirrander}, \citenamefont {Robinson}, \citenamefont {Ratner}, \citenamefont {Lane}, \citenamefont {Zhu}, \citenamefont {Glownia}, \citenamefont {Kozina}, \citenamefont {Lemke}, \citenamefont {Sikorski}, \citenamefont {Feng}, \citenamefont {Nelson}, \citenamefont {Saita}, \citenamefont {Stankus}, \citenamefont {Northey}, \citenamefont {Hastings},\ and\ \citenamefont {Weber}}]{CHD266LCLS}%
  \BibitemOpen
  \bibfield  {author} {\bibinfo {author} {\bibfnamefont {M.~P.}\ \bibnamefont {Minitti}}, \bibinfo {author} {\bibfnamefont {J.~M.}\ \bibnamefont {Budarz}}, \bibinfo {author} {\bibfnamefont {A.}~\bibnamefont {Kirrander}}, \bibinfo {author} {\bibfnamefont {J.~S.}\ \bibnamefont {Robinson}}, \bibinfo {author} {\bibfnamefont {D.}~\bibnamefont {Ratner}}, \bibinfo {author} {\bibfnamefont {T.~J.}\ \bibnamefont {Lane}}, \bibinfo {author} {\bibfnamefont {D.}~\bibnamefont {Zhu}}, \bibinfo {author} {\bibfnamefont {J.~M.}\ \bibnamefont {Glownia}}, \bibinfo {author} {\bibfnamefont {M.}~\bibnamefont {Kozina}}, \bibinfo {author} {\bibfnamefont {H.~T.}\ \bibnamefont {Lemke}}, \bibinfo {author} {\bibfnamefont {M.}~\bibnamefont {Sikorski}}, \bibinfo {author} {\bibfnamefont {Y.}~\bibnamefont {Feng}}, \bibinfo {author} {\bibfnamefont {S.}~\bibnamefont {Nelson}}, \bibinfo {author} {\bibfnamefont {K.}~\bibnamefont {Saita}}, \bibinfo {author} {\bibfnamefont {B.}~\bibnamefont {Stankus}}, \bibinfo {author} {\bibfnamefont
  {T.}~\bibnamefont {Northey}}, \bibinfo {author} {\bibfnamefont {J.~B.}\ \bibnamefont {Hastings}},\ and\ \bibinfo {author} {\bibfnamefont {P.~M.}\ \bibnamefont {Weber}},\ }\bibfield  {title} {\bibinfo {title} {Imaging molecular motion: Femtosecond x-ray scattering of an electrocyclic chemical reaction},\ }\href {https://doi.org/10.1103/PhysRevLett.114.255501} {\bibfield  {journal} {\bibinfo  {journal} {Phys. Rev. Lett.}\ }\textbf {\bibinfo {volume} {114}},\ \bibinfo {pages} {255501} (\bibinfo {year} {2015})}\BibitemShut {NoStop}%
\bibitem [{\citenamefont {Glownia}\ \emph {et~al.}(2016)\citenamefont {Glownia}, \citenamefont {Natan}, \citenamefont {Cryan}, \citenamefont {Hartsock}, \citenamefont {Kozina}, \citenamefont {Minitti}, \citenamefont {Nelson}, \citenamefont {Robinson}, \citenamefont {Sato}, \citenamefont {van Driel}, \citenamefont {Welch}, \citenamefont {Weninger}, \citenamefont {Zhu},\ and\ \citenamefont {Bucksbaum}}]{I2LCLS}%
  \BibitemOpen
  \bibfield  {author} {\bibinfo {author} {\bibfnamefont {J.~M.}\ \bibnamefont {Glownia}}, \bibinfo {author} {\bibfnamefont {A.}~\bibnamefont {Natan}}, \bibinfo {author} {\bibfnamefont {J.~P.}\ \bibnamefont {Cryan}}, \bibinfo {author} {\bibfnamefont {R.}~\bibnamefont {Hartsock}}, \bibinfo {author} {\bibfnamefont {M.}~\bibnamefont {Kozina}}, \bibinfo {author} {\bibfnamefont {M.~P.}\ \bibnamefont {Minitti}}, \bibinfo {author} {\bibfnamefont {S.}~\bibnamefont {Nelson}}, \bibinfo {author} {\bibfnamefont {J.}~\bibnamefont {Robinson}}, \bibinfo {author} {\bibfnamefont {T.}~\bibnamefont {Sato}}, \bibinfo {author} {\bibfnamefont {T.}~\bibnamefont {van Driel}}, \bibinfo {author} {\bibfnamefont {G.}~\bibnamefont {Welch}}, \bibinfo {author} {\bibfnamefont {C.}~\bibnamefont {Weninger}}, \bibinfo {author} {\bibfnamefont {D.}~\bibnamefont {Zhu}},\ and\ \bibinfo {author} {\bibfnamefont {P.~H.}\ \bibnamefont {Bucksbaum}},\ }\bibfield  {title} {\bibinfo {title} {Self-referenced coherent diffraction x-ray movie of
  {\aa{}}ngstrom- and femtosecond-scale atomic motion},\ }\href {https://doi.org/10.1103/PhysRevLett.117.153003} {\bibfield  {journal} {\bibinfo  {journal} {Phys. Rev. Lett.}\ }\textbf {\bibinfo {volume} {117}},\ \bibinfo {pages} {153003} (\bibinfo {year} {2016})}\BibitemShut {NoStop}%
\bibitem [{\citenamefont {Stankus}\ \emph {et~al.}(2019)\citenamefont {Stankus}, \citenamefont {Yong}, \citenamefont {Zotev}, \citenamefont {Ruddock}, \citenamefont {Bellshaw}, \citenamefont {Lane}, \citenamefont {Liang}, \citenamefont {Boutet}, \citenamefont {Carbajo}, \citenamefont {Robinson}, \citenamefont {Du}, \citenamefont {Goff}, \citenamefont {Chang}, \citenamefont {Koglin}, \citenamefont {Minitti}, \citenamefont {Kirrander},\ and\ \citenamefont {Weber}}]{stankus2019}%
  \BibitemOpen
  \bibfield  {author} {\bibinfo {author} {\bibfnamefont {B.}~\bibnamefont {Stankus}}, \bibinfo {author} {\bibfnamefont {H.}~\bibnamefont {Yong}}, \bibinfo {author} {\bibfnamefont {N.}~\bibnamefont {Zotev}}, \bibinfo {author} {\bibfnamefont {J.~M.}\ \bibnamefont {Ruddock}}, \bibinfo {author} {\bibfnamefont {D.}~\bibnamefont {Bellshaw}}, \bibinfo {author} {\bibfnamefont {T.~J.}\ \bibnamefont {Lane}}, \bibinfo {author} {\bibfnamefont {M.}~\bibnamefont {Liang}}, \bibinfo {author} {\bibfnamefont {S.}~\bibnamefont {Boutet}}, \bibinfo {author} {\bibfnamefont {S.}~\bibnamefont {Carbajo}}, \bibinfo {author} {\bibfnamefont {J.~S.}\ \bibnamefont {Robinson}}, \bibinfo {author} {\bibfnamefont {W.}~\bibnamefont {Du}}, \bibinfo {author} {\bibfnamefont {N.}~\bibnamefont {Goff}}, \bibinfo {author} {\bibfnamefont {Y.}~\bibnamefont {Chang}}, \bibinfo {author} {\bibfnamefont {J.~E.}\ \bibnamefont {Koglin}}, \bibinfo {author} {\bibfnamefont {M.~P.}\ \bibnamefont {Minitti}}, \bibinfo {author} {\bibfnamefont {A.}~\bibnamefont
  {Kirrander}},\ and\ \bibinfo {author} {\bibfnamefont {P.~M.}\ \bibnamefont {Weber}},\ }\bibfield  {title} {\bibinfo {title} {Ultrafast {X}-ray scattering reveals vibrational coherence following {Rydberg} excitation},\ }\href {https://doi.org/10.1038/s41557-019-0291-0} {\bibfield  {journal} {\bibinfo  {journal} {Nature Chem.}\ }\textbf {\bibinfo {volume} {11}},\ \bibinfo {pages} {716} (\bibinfo {year} {2019})}\BibitemShut {NoStop}%
\bibitem [{\citenamefont {Ruddock}\ \emph {et~al.}(2019)\citenamefont {Ruddock}, \citenamefont {Yong}, \citenamefont {Stankus}, \citenamefont {Du}, \citenamefont {Goff}, \citenamefont {Chang}, \citenamefont {Odate}, \citenamefont {Carrascosa}, \citenamefont {Bellshaw}, \citenamefont {Zotev}, \citenamefont {Liang}, \citenamefont {Carbajo}, \citenamefont {Koglin}, \citenamefont {Robinson}, \citenamefont {Boutet}, \citenamefont {Kirrander}, \citenamefont {Minitti},\ and\ \citenamefont {Weber}}]{ruddock_deep_2019}%
  \BibitemOpen
  \bibfield  {author} {\bibinfo {author} {\bibfnamefont {J.~M.}\ \bibnamefont {Ruddock}}, \bibinfo {author} {\bibfnamefont {H.}~\bibnamefont {Yong}}, \bibinfo {author} {\bibfnamefont {B.}~\bibnamefont {Stankus}}, \bibinfo {author} {\bibfnamefont {W.}~\bibnamefont {Du}}, \bibinfo {author} {\bibfnamefont {N.}~\bibnamefont {Goff}}, \bibinfo {author} {\bibfnamefont {Y.}~\bibnamefont {Chang}}, \bibinfo {author} {\bibfnamefont {A.}~\bibnamefont {Odate}}, \bibinfo {author} {\bibfnamefont {A.~M.}\ \bibnamefont {Carrascosa}}, \bibinfo {author} {\bibfnamefont {D.}~\bibnamefont {Bellshaw}}, \bibinfo {author} {\bibfnamefont {N.}~\bibnamefont {Zotev}}, \bibinfo {author} {\bibfnamefont {M.}~\bibnamefont {Liang}}, \bibinfo {author} {\bibfnamefont {S.}~\bibnamefont {Carbajo}}, \bibinfo {author} {\bibfnamefont {J.}~\bibnamefont {Koglin}}, \bibinfo {author} {\bibfnamefont {J.~S.}\ \bibnamefont {Robinson}}, \bibinfo {author} {\bibfnamefont {S.}~\bibnamefont {Boutet}}, \bibinfo {author} {\bibfnamefont {A.}~\bibnamefont
  {Kirrander}}, \bibinfo {author} {\bibfnamefont {M.~P.}\ \bibnamefont {Minitti}},\ and\ \bibinfo {author} {\bibfnamefont {P.~M.}\ \bibnamefont {Weber}},\ }\bibfield  {title} {\bibinfo {title} {A deep {UV} trigger for ground-state ring-opening dynamics of 1,3-cyclohexadiene},\ }\href {https://doi.org/10.1126/sciadv.aax6625} {\bibfield  {journal} {\bibinfo  {journal} {Science Advances}\ }\textbf {\bibinfo {volume} {5}},\ \bibinfo {pages} {eaax6625} (\bibinfo {year} {2019})}\BibitemShut {NoStop}%
\bibitem [{\citenamefont {Yong}\ \emph {et~al.}(2020)\citenamefont {Yong}, \citenamefont {Zotev}, \citenamefont {Ruddock}, \citenamefont {Stankus}, \citenamefont {Simmermacher}, \citenamefont {Carrascosa}, \citenamefont {Du}, \citenamefont {Goff}, \citenamefont {Chang}, \citenamefont {Bellshaw}, \citenamefont {Liang}, \citenamefont {Carbajo}, \citenamefont {Koglin}, \citenamefont {Robinson}, \citenamefont {Boutet}, \citenamefont {Minitti}, \citenamefont {Kirrander},\ and\ \citenamefont {Weber}}]{yong_observation_2020}%
  \BibitemOpen
  \bibfield  {author} {\bibinfo {author} {\bibfnamefont {H.}~\bibnamefont {Yong}}, \bibinfo {author} {\bibfnamefont {N.}~\bibnamefont {Zotev}}, \bibinfo {author} {\bibfnamefont {J.~M.}\ \bibnamefont {Ruddock}}, \bibinfo {author} {\bibfnamefont {B.}~\bibnamefont {Stankus}}, \bibinfo {author} {\bibfnamefont {M.}~\bibnamefont {Simmermacher}}, \bibinfo {author} {\bibfnamefont {A.~M.}\ \bibnamefont {Carrascosa}}, \bibinfo {author} {\bibfnamefont {W.}~\bibnamefont {Du}}, \bibinfo {author} {\bibfnamefont {N.}~\bibnamefont {Goff}}, \bibinfo {author} {\bibfnamefont {Y.}~\bibnamefont {Chang}}, \bibinfo {author} {\bibfnamefont {D.}~\bibnamefont {Bellshaw}}, \bibinfo {author} {\bibfnamefont {M.}~\bibnamefont {Liang}}, \bibinfo {author} {\bibfnamefont {S.}~\bibnamefont {Carbajo}}, \bibinfo {author} {\bibfnamefont {J.~E.}\ \bibnamefont {Koglin}}, \bibinfo {author} {\bibfnamefont {J.~S.}\ \bibnamefont {Robinson}}, \bibinfo {author} {\bibfnamefont {S.}~\bibnamefont {Boutet}}, \bibinfo {author} {\bibfnamefont {M.~P.}\
  \bibnamefont {Minitti}}, \bibinfo {author} {\bibfnamefont {A.}~\bibnamefont {Kirrander}},\ and\ \bibinfo {author} {\bibfnamefont {P.~M.}\ \bibnamefont {Weber}},\ }\bibfield  {title} {\bibinfo {title} {Observation of the molecular response to light upon photoexcitation},\ }\href {https://doi.org/10.1038/s41467-020-15680-4} {\bibfield  {journal} {\bibinfo  {journal} {Nature Commun.}\ }\textbf {\bibinfo {volume} {11}},\ \bibinfo {pages} {2157} (\bibinfo {year} {2020})}\BibitemShut {NoStop}%
\bibitem [{\citenamefont {Yong}\ \emph {et~al.}(2021)\citenamefont {Yong}, \citenamefont {Xu}, \citenamefont {Ruddock}, \citenamefont {Stankus}, \citenamefont {Carrascosa}, \citenamefont {Zotev}, \citenamefont {Bellshaw}, \citenamefont {Du}, \citenamefont {Goff}, \citenamefont {Chang}, \citenamefont {Boutet}, \citenamefont {Carbajo}, \citenamefont {Koglin}, \citenamefont {Liang}, \citenamefont {Robinson}, \citenamefont {Kirrander}, \citenamefont {Minitti},\ and\ \citenamefont {Weber}}]{yong_ultrafast_2021}%
  \BibitemOpen
  \bibfield  {author} {\bibinfo {author} {\bibfnamefont {H.}~\bibnamefont {Yong}}, \bibinfo {author} {\bibfnamefont {X.}~\bibnamefont {Xu}}, \bibinfo {author} {\bibfnamefont {J.~M.}\ \bibnamefont {Ruddock}}, \bibinfo {author} {\bibfnamefont {B.}~\bibnamefont {Stankus}}, \bibinfo {author} {\bibfnamefont {A.~M.}\ \bibnamefont {Carrascosa}}, \bibinfo {author} {\bibfnamefont {N.}~\bibnamefont {Zotev}}, \bibinfo {author} {\bibfnamefont {D.}~\bibnamefont {Bellshaw}}, \bibinfo {author} {\bibfnamefont {W.}~\bibnamefont {Du}}, \bibinfo {author} {\bibfnamefont {N.}~\bibnamefont {Goff}}, \bibinfo {author} {\bibfnamefont {Y.}~\bibnamefont {Chang}}, \bibinfo {author} {\bibfnamefont {S.}~\bibnamefont {Boutet}}, \bibinfo {author} {\bibfnamefont {S.}~\bibnamefont {Carbajo}}, \bibinfo {author} {\bibfnamefont {J.~E.}\ \bibnamefont {Koglin}}, \bibinfo {author} {\bibfnamefont {M.}~\bibnamefont {Liang}}, \bibinfo {author} {\bibfnamefont {J.~S.}\ \bibnamefont {Robinson}}, \bibinfo {author} {\bibfnamefont {A.}~\bibnamefont
  {Kirrander}}, \bibinfo {author} {\bibfnamefont {M.~P.}\ \bibnamefont {Minitti}},\ and\ \bibinfo {author} {\bibfnamefont {P.~M.}\ \bibnamefont {Weber}},\ }\bibfield  {title} {\bibinfo {title} {Ultrafast {X}-ray scattering offers a structural view of excited-state charge transfer},\ }\href {https://doi.org/10.1073/pnas.2021714118} {\bibfield  {journal} {\bibinfo  {journal} {Proc. Natl. Acad. Sci.}\ }\textbf {\bibinfo {volume} {118}},\ \bibinfo {pages} {e2021714118} (\bibinfo {year} {2021})}\BibitemShut {NoStop}%
\bibitem [{\citenamefont {Yang}\ \emph {et~al.}(2016)\citenamefont {Yang}, \citenamefont {Guehr}, \citenamefont {Shen}, \citenamefont {Li}, \citenamefont {Vecchione}, \citenamefont {Coffee}, \citenamefont {Corbett}, \citenamefont {Fry}, \citenamefont {Hartmann}, \citenamefont {Hast}, \citenamefont {Hegazy}, \citenamefont {Jobe}, \citenamefont {Makasyuk}, \citenamefont {Robinson}, \citenamefont {Robinson}, \citenamefont {Vetter}, \citenamefont {Weathersby}, \citenamefont {Yoneda}, \citenamefont {Wang},\ and\ \citenamefont {Centurion}}]{YangJ16:PRL117}%
  \BibitemOpen
  \bibfield  {author} {\bibinfo {author} {\bibfnamefont {J.}~\bibnamefont {Yang}}, \bibinfo {author} {\bibfnamefont {M.}~\bibnamefont {Guehr}}, \bibinfo {author} {\bibfnamefont {X.}~\bibnamefont {Shen}}, \bibinfo {author} {\bibfnamefont {R.}~\bibnamefont {Li}}, \bibinfo {author} {\bibfnamefont {T.}~\bibnamefont {Vecchione}}, \bibinfo {author} {\bibfnamefont {R.}~\bibnamefont {Coffee}}, \bibinfo {author} {\bibfnamefont {J.}~\bibnamefont {Corbett}}, \bibinfo {author} {\bibfnamefont {A.}~\bibnamefont {Fry}}, \bibinfo {author} {\bibfnamefont {N.}~\bibnamefont {Hartmann}}, \bibinfo {author} {\bibfnamefont {C.}~\bibnamefont {Hast}}, \bibinfo {author} {\bibfnamefont {K.}~\bibnamefont {Hegazy}}, \bibinfo {author} {\bibfnamefont {K.}~\bibnamefont {Jobe}}, \bibinfo {author} {\bibfnamefont {I.}~\bibnamefont {Makasyuk}}, \bibinfo {author} {\bibfnamefont {J.}~\bibnamefont {Robinson}}, \bibinfo {author} {\bibfnamefont {M.~S.}\ \bibnamefont {Robinson}}, \bibinfo {author} {\bibfnamefont {S.}~\bibnamefont {Vetter}}, \bibinfo
  {author} {\bibfnamefont {S.}~\bibnamefont {Weathersby}}, \bibinfo {author} {\bibfnamefont {C.}~\bibnamefont {Yoneda}}, \bibinfo {author} {\bibfnamefont {X.}~\bibnamefont {Wang}},\ and\ \bibinfo {author} {\bibfnamefont {M.}~\bibnamefont {Centurion}},\ }\bibfield  {title} {\bibinfo {title} {Diffractive imaging of coherent nuclear motion in isolated molecules},\ }\href@noop {} {\bibfield  {journal} {\bibinfo  {journal} {Phys. Rev. Lett.}\ }\textbf {\bibinfo {volume} {117}},\ \bibinfo {pages} {153002} (\bibinfo {year} {2016})}\BibitemShut {NoStop}%
\bibitem [{\citenamefont {Yang}\ \emph {et~al.}(2018)\citenamefont {Yang}, \citenamefont {Zhu}, \citenamefont {Wolf}, \citenamefont {Li}, \citenamefont {Nunes}, \citenamefont {Coffee}, \citenamefont {Cryan}, \citenamefont {Gühr}, \citenamefont {Hegazy}, \citenamefont {Heinz}, \citenamefont {Jobe}, \citenamefont {Li}, \citenamefont {Shen}, \citenamefont {Veccione}, \citenamefont {Weathersby}, \citenamefont {Wilkin}, \citenamefont {Yoneda}, \citenamefont {Zheng}, \citenamefont {Martinez}, \citenamefont {Centurion},\ and\ \citenamefont {Wang}}]{YangJ18:Scicen361}%
  \BibitemOpen
  \bibfield  {author} {\bibinfo {author} {\bibfnamefont {J.}~\bibnamefont {Yang}}, \bibinfo {author} {\bibfnamefont {X.}~\bibnamefont {Zhu}}, \bibinfo {author} {\bibfnamefont {T.~J.~A.}\ \bibnamefont {Wolf}}, \bibinfo {author} {\bibfnamefont {Z.}~\bibnamefont {Li}}, \bibinfo {author} {\bibfnamefont {J.~P.~F.}\ \bibnamefont {Nunes}}, \bibinfo {author} {\bibfnamefont {R.}~\bibnamefont {Coffee}}, \bibinfo {author} {\bibfnamefont {J.~P.}\ \bibnamefont {Cryan}}, \bibinfo {author} {\bibfnamefont {M.}~\bibnamefont {Gühr}}, \bibinfo {author} {\bibfnamefont {K.}~\bibnamefont {Hegazy}}, \bibinfo {author} {\bibfnamefont {T.~F.}\ \bibnamefont {Heinz}}, \bibinfo {author} {\bibfnamefont {K.}~\bibnamefont {Jobe}}, \bibinfo {author} {\bibfnamefont {R.}~\bibnamefont {Li}}, \bibinfo {author} {\bibfnamefont {X.}~\bibnamefont {Shen}}, \bibinfo {author} {\bibfnamefont {T.}~\bibnamefont {Veccione}}, \bibinfo {author} {\bibfnamefont {S.}~\bibnamefont {Weathersby}}, \bibinfo {author} {\bibfnamefont {K.~J.}\ \bibnamefont {Wilkin}},
  \bibinfo {author} {\bibfnamefont {C.}~\bibnamefont {Yoneda}}, \bibinfo {author} {\bibfnamefont {Q.}~\bibnamefont {Zheng}}, \bibinfo {author} {\bibfnamefont {T.~J.}\ \bibnamefont {Martinez}}, \bibinfo {author} {\bibfnamefont {M.}~\bibnamefont {Centurion}},\ and\ \bibinfo {author} {\bibfnamefont {X.}~\bibnamefont {Wang}},\ }\bibfield  {title} {\bibinfo {title} {Imaging {CF$_3$I} conical intersection and photodissociation dynamics with ultrafast electron diffraction},\ }\href@noop {} {\bibfield  {journal} {\bibinfo  {journal} {Science}\ }\textbf {\bibinfo {volume} {361}},\ \bibinfo {pages} {64} (\bibinfo {year} {2018})}\BibitemShut {NoStop}%
\bibitem [{\citenamefont {Wolf}\ \emph {et~al.}(2019)\citenamefont {Wolf}, \citenamefont {Sanchez}, \citenamefont {Yang}, \citenamefont {Parrish}, \citenamefont {Nunes}, \citenamefont {Centurion}, \citenamefont {Coffee}, \citenamefont {Cryan}, \citenamefont {G\"uhr}, \citenamefont {Hegazy}, \citenamefont {Kirrander}, \citenamefont {Li}, \citenamefont {Ruddock}, \citenamefont {Shen}, \citenamefont {Vecchione}, \citenamefont {Weathersby}, \citenamefont {Weber}, \citenamefont {Wilkin}, \citenamefont {Yong}, \citenamefont {Zheng}, \citenamefont {Wang}, \citenamefont {Minitti},\ and\ \citenamefont {Mart\'inez}}]{Wolf19:NatChem504}%
  \BibitemOpen
  \bibfield  {author} {\bibinfo {author} {\bibfnamefont {T.}~\bibnamefont {Wolf}}, \bibinfo {author} {\bibfnamefont {D.~M.}\ \bibnamefont {Sanchez}}, \bibinfo {author} {\bibfnamefont {J.}~\bibnamefont {Yang}}, \bibinfo {author} {\bibfnamefont {R.}~\bibnamefont {Parrish}}, \bibinfo {author} {\bibfnamefont {J.}~\bibnamefont {Nunes}}, \bibinfo {author} {\bibfnamefont {M.}~\bibnamefont {Centurion}}, \bibinfo {author} {\bibfnamefont {R.}~\bibnamefont {Coffee}}, \bibinfo {author} {\bibfnamefont {J.}~\bibnamefont {Cryan}}, \bibinfo {author} {\bibfnamefont {M.}~\bibnamefont {G\"uhr}}, \bibinfo {author} {\bibfnamefont {K.}~\bibnamefont {Hegazy}}, \bibinfo {author} {\bibfnamefont {A.}~\bibnamefont {Kirrander}}, \bibinfo {author} {\bibfnamefont {R.~K.}\ \bibnamefont {Li}}, \bibinfo {author} {\bibfnamefont {J.}~\bibnamefont {Ruddock}}, \bibinfo {author} {\bibfnamefont {X.}~\bibnamefont {Shen}}, \bibinfo {author} {\bibfnamefont {T.}~\bibnamefont {Vecchione}}, \bibinfo {author} {\bibfnamefont {S.~P.}\ \bibnamefont
  {Weathersby}}, \bibinfo {author} {\bibfnamefont {P.~M.}\ \bibnamefont {Weber}}, \bibinfo {author} {\bibfnamefont {K.}~\bibnamefont {Wilkin}}, \bibinfo {author} {\bibfnamefont {H.}~\bibnamefont {Yong}}, \bibinfo {author} {\bibfnamefont {Q.}~\bibnamefont {Zheng}}, \bibinfo {author} {\bibfnamefont {X.~J.}\ \bibnamefont {Wang}}, \bibinfo {author} {\bibfnamefont {M.~P.}\ \bibnamefont {Minitti}},\ and\ \bibinfo {author} {\bibfnamefont {T.~J.}\ \bibnamefont {Mart\'inez}},\ }\bibfield  {title} {\bibinfo {title} {The photochemical ring-opening of 1,3-cyclohexadiene imaged by ultrafast electron diffraction},\ }\href@noop {} {\bibfield  {journal} {\bibinfo  {journal} {Nature Chem.}\ }\textbf {\bibinfo {volume} {11}},\ \bibinfo {pages} {504} (\bibinfo {year} {2019})}\BibitemShut {NoStop}%
\bibitem [{\citenamefont {Yang}\ \emph {et~al.}(2020)\citenamefont {Yang}, \citenamefont {Zhu}, \citenamefont {F.~Nunes}, \citenamefont {Yu}, \citenamefont {Parrish}, \citenamefont {Wolf}, \citenamefont {Centurion}, \citenamefont {Gühr}, \citenamefont {Li}, \citenamefont {Liu}, \citenamefont {Moore}, \citenamefont {Niebuhr}, \citenamefont {Park}, \citenamefont {Shen}, \citenamefont {Weathersby}, \citenamefont {Weinacht}, \citenamefont {Martinez},\ and\ \citenamefont {Wang}}]{YangJ20:Science368}%
  \BibitemOpen
  \bibfield  {author} {\bibinfo {author} {\bibfnamefont {J.}~\bibnamefont {Yang}}, \bibinfo {author} {\bibfnamefont {X.}~\bibnamefont {Zhu}}, \bibinfo {author} {\bibfnamefont {J.~P.}\ \bibnamefont {F.~Nunes}}, \bibinfo {author} {\bibfnamefont {J.~K.}\ \bibnamefont {Yu}}, \bibinfo {author} {\bibfnamefont {R.~M.}\ \bibnamefont {Parrish}}, \bibinfo {author} {\bibfnamefont {T.~J.~A.}\ \bibnamefont {Wolf}}, \bibinfo {author} {\bibfnamefont {M.}~\bibnamefont {Centurion}}, \bibinfo {author} {\bibfnamefont {M.}~\bibnamefont {Gühr}}, \bibinfo {author} {\bibfnamefont {R.}~\bibnamefont {Li}}, \bibinfo {author} {\bibfnamefont {Y.}~\bibnamefont {Liu}}, \bibinfo {author} {\bibfnamefont {B.}~\bibnamefont {Moore}}, \bibinfo {author} {\bibfnamefont {M.}~\bibnamefont {Niebuhr}}, \bibinfo {author} {\bibfnamefont {S.}~\bibnamefont {Park}}, \bibinfo {author} {\bibfnamefont {X.}~\bibnamefont {Shen}}, \bibinfo {author} {\bibfnamefont {S.}~\bibnamefont {Weathersby}}, \bibinfo {author} {\bibfnamefont {T.}~\bibnamefont {Weinacht}},
  \bibinfo {author} {\bibfnamefont {T.~J.}\ \bibnamefont {Martinez}},\ and\ \bibinfo {author} {\bibfnamefont {X.}~\bibnamefont {Wang}},\ }\bibfield  {title} {\bibinfo {title} {Simultaneous observation of nuclear and electronic dynamics by ultrafast electron diffraction},\ }\href@noop {} {\bibfield  {journal} {\bibinfo  {journal} {Science}\ }\textbf {\bibinfo {volume} {368}},\ \bibinfo {pages} {885} (\bibinfo {year} {2020})}\BibitemShut {NoStop}%
\bibitem [{\citenamefont {Wang}\ \emph {et~al.}(2025)\citenamefont {Wang}, \citenamefont {Jiang}, \citenamefont {Jin}, \citenamefont {Zou}, \citenamefont {Zhu}, \citenamefont {Jiang}, \citenamefont {He},\ and\ \citenamefont {Xiang}}]{wang_imaging_2025}%
  \BibitemOpen
  \bibfield  {author} {\bibinfo {author} {\bibfnamefont {T.}~\bibnamefont {Wang}}, \bibinfo {author} {\bibfnamefont {H.}~\bibnamefont {Jiang}}, \bibinfo {author} {\bibfnamefont {C.}~\bibnamefont {Jin}}, \bibinfo {author} {\bibfnamefont {X.}~\bibnamefont {Zou}}, \bibinfo {author} {\bibfnamefont {P.}~\bibnamefont {Zhu}}, \bibinfo {author} {\bibfnamefont {T.}~\bibnamefont {Jiang}}, \bibinfo {author} {\bibfnamefont {F.}~\bibnamefont {He}},\ and\ \bibinfo {author} {\bibfnamefont {D.}~\bibnamefont {Xiang}},\ }\bibfield  {title} {\bibinfo {title} {Imaging the photochemical dynamics of cyclobutanone with {MeV} ultrafast electron diffraction},\ }\href {https://doi.org/10.1063/5.0267186} {\bibfield  {journal} {\bibinfo  {journal} {J. Chem. Phys.}\ }\textbf {\bibinfo {volume} {162}},\ \bibinfo {pages} {184201} (\bibinfo {year} {2025})}\BibitemShut {NoStop}%
\bibitem [{\citenamefont {Green}\ \emph {et~al.}(2025)\citenamefont {Green}, \citenamefont {Liu}, \citenamefont {Allum}, \citenamefont {Graßl}, \citenamefont {Lenzen}, \citenamefont {Ashfold}, \citenamefont {Bhattacharyya}, \citenamefont {Cheng}, \citenamefont {Centurion}, \citenamefont {Crane}, \citenamefont {Forbes}, \citenamefont {Goff}, \citenamefont {Huang}, \citenamefont {Kaufman}, \citenamefont {Kling}, \citenamefont {Kramer}, \citenamefont {Lam}, \citenamefont {Larsen}, \citenamefont {Lemons}, \citenamefont {Lin}, \citenamefont {Orr-Ewing}, \citenamefont {Rolles}, \citenamefont {Rudenko}, \citenamefont {Saha}, \citenamefont {Searles}, \citenamefont {Shen}, \citenamefont {Weathersby}, \citenamefont {Weber}, \citenamefont {Zhao},\ and\ \citenamefont {Wolf}}]{green_imaging_2025}%
  \BibitemOpen
  \bibfield  {author} {\bibinfo {author} {\bibfnamefont {A.~E.}\ \bibnamefont {Green}}, \bibinfo {author} {\bibfnamefont {Y.}~\bibnamefont {Liu}}, \bibinfo {author} {\bibfnamefont {F.}~\bibnamefont {Allum}}, \bibinfo {author} {\bibfnamefont {M.}~\bibnamefont {Graßl}}, \bibinfo {author} {\bibfnamefont {P.}~\bibnamefont {Lenzen}}, \bibinfo {author} {\bibfnamefont {M.~N.~R.}\ \bibnamefont {Ashfold}}, \bibinfo {author} {\bibfnamefont {S.}~\bibnamefont {Bhattacharyya}}, \bibinfo {author} {\bibfnamefont {X.}~\bibnamefont {Cheng}}, \bibinfo {author} {\bibfnamefont {M.}~\bibnamefont {Centurion}}, \bibinfo {author} {\bibfnamefont {S.~W.}\ \bibnamefont {Crane}}, \bibinfo {author} {\bibfnamefont {R.}~\bibnamefont {Forbes}}, \bibinfo {author} {\bibfnamefont {N.~A.}\ \bibnamefont {Goff}}, \bibinfo {author} {\bibfnamefont {L.}~\bibnamefont {Huang}}, \bibinfo {author} {\bibfnamefont {B.}~\bibnamefont {Kaufman}}, \bibinfo {author} {\bibfnamefont {M.-F.}\ \bibnamefont {Kling}}, \bibinfo {author} {\bibfnamefont {P.~L.}\
  \bibnamefont {Kramer}}, \bibinfo {author} {\bibfnamefont {H.~V.~S.}\ \bibnamefont {Lam}}, \bibinfo {author} {\bibfnamefont {K.~A.}\ \bibnamefont {Larsen}}, \bibinfo {author} {\bibfnamefont {R.}~\bibnamefont {Lemons}}, \bibinfo {author} {\bibfnamefont {M.-F.}\ \bibnamefont {Lin}}, \bibinfo {author} {\bibfnamefont {A.~J.}\ \bibnamefont {Orr-Ewing}}, \bibinfo {author} {\bibfnamefont {D.}~\bibnamefont {Rolles}}, \bibinfo {author} {\bibfnamefont {A.}~\bibnamefont {Rudenko}}, \bibinfo {author} {\bibfnamefont {S.~K.}\ \bibnamefont {Saha}}, \bibinfo {author} {\bibfnamefont {J.}~\bibnamefont {Searles}}, \bibinfo {author} {\bibfnamefont {X.}~\bibnamefont {Shen}}, \bibinfo {author} {\bibfnamefont {S.}~\bibnamefont {Weathersby}}, \bibinfo {author} {\bibfnamefont {P.~M.}\ \bibnamefont {Weber}}, \bibinfo {author} {\bibfnamefont {H.}~\bibnamefont {Zhao}},\ and\ \bibinfo {author} {\bibfnamefont {T.~J.~A.}\ \bibnamefont {Wolf}},\ }\bibfield  {title} {\bibinfo {title} {Imaging the photochemistry of cyclobutanone using
  ultrafast electron diffraction: {Experimental} results},\ }\href {https://doi.org/10.1063/5.0266559} {\bibfield  {journal} {\bibinfo  {journal} {J. Chem. Phys.}\ }\textbf {\bibinfo {volume} {162}},\ \bibinfo {pages} {184303} (\bibinfo {year} {2025})}\BibitemShut {NoStop}%
\bibitem [{\citenamefont {Champenois}\ \emph {et~al.}(2023)\citenamefont {Champenois}, \citenamefont {List}, \citenamefont {Ware}, \citenamefont {Britton}, \citenamefont {Bucksbaum}, \citenamefont {Cheng}, \citenamefont {Centurion}, \citenamefont {Cryan}, \citenamefont {Forbes}, \citenamefont {Gabalski}, \citenamefont {Hegazy}, \citenamefont {Hoffmann}, \citenamefont {Howard}, \citenamefont {Ji}, \citenamefont {Lin}, \citenamefont {Nunes}, \citenamefont {Shen}, \citenamefont {Yang}, \citenamefont {Wang}, \citenamefont {Martinez},\ and\ \citenamefont {Wolf}}]{ThomasWolf131:PRL23}%
  \BibitemOpen
  \bibfield  {author} {\bibinfo {author} {\bibfnamefont {E.~G.}\ \bibnamefont {Champenois}}, \bibinfo {author} {\bibfnamefont {N.~H.}\ \bibnamefont {List}}, \bibinfo {author} {\bibfnamefont {M.}~\bibnamefont {Ware}}, \bibinfo {author} {\bibfnamefont {M.}~\bibnamefont {Britton}}, \bibinfo {author} {\bibfnamefont {P.~H.}\ \bibnamefont {Bucksbaum}}, \bibinfo {author} {\bibfnamefont {X.}~\bibnamefont {Cheng}}, \bibinfo {author} {\bibfnamefont {M.}~\bibnamefont {Centurion}}, \bibinfo {author} {\bibfnamefont {J.~P.}\ \bibnamefont {Cryan}}, \bibinfo {author} {\bibfnamefont {R.}~\bibnamefont {Forbes}}, \bibinfo {author} {\bibfnamefont {I.}~\bibnamefont {Gabalski}}, \bibinfo {author} {\bibfnamefont {K.}~\bibnamefont {Hegazy}}, \bibinfo {author} {\bibfnamefont {M.~C.}\ \bibnamefont {Hoffmann}}, \bibinfo {author} {\bibfnamefont {A.~J.}\ \bibnamefont {Howard}}, \bibinfo {author} {\bibfnamefont {F.}~\bibnamefont {Ji}}, \bibinfo {author} {\bibfnamefont {M.-F.}\ \bibnamefont {Lin}}, \bibinfo {author} {\bibfnamefont
  {J.~P.~F.}\ \bibnamefont {Nunes}}, \bibinfo {author} {\bibfnamefont {X.}~\bibnamefont {Shen}}, \bibinfo {author} {\bibfnamefont {J.}~\bibnamefont {Yang}}, \bibinfo {author} {\bibfnamefont {X.}~\bibnamefont {Wang}}, \bibinfo {author} {\bibfnamefont {T.~J.}\ \bibnamefont {Martinez}},\ and\ \bibinfo {author} {\bibfnamefont {T.~J.~A.}\ \bibnamefont {Wolf}},\ }\bibfield  {title} {\bibinfo {title} {Femtosecond electronic and hydrogen structural dynamics in ammonia imaged with ultrafast electron diffraction},\ }\href@noop {} {\bibfield  {journal} {\bibinfo  {journal} {Phys. Rev. Lett.}\ }\textbf {\bibinfo {volume} {131}},\ \bibinfo {pages} {143001} (\bibinfo {year} {2023})}\BibitemShut {NoStop}%
\bibitem [{\citenamefont {Zhu}\ and\ \citenamefont {Yarkony}(2012)}]{Zhu137:JCP12}%
  \BibitemOpen
  \bibfield  {author} {\bibinfo {author} {\bibfnamefont {X.}~\bibnamefont {Zhu}}\ and\ \bibinfo {author} {\bibfnamefont {D.~R.}\ \bibnamefont {Yarkony}},\ }\bibfield  {title} {\bibinfo {title} {Quasi-diabatic representations of adiabatic potential energy surfaces coupled by conical intersections including bond breaking: A more general construction procedure and an analysis of the diabatic representation},\ }\href@noop {} {\bibfield  {journal} {\bibinfo  {journal} {J. Chem. Phys.}\ }\textbf {\bibinfo {volume} {137}},\ \bibinfo {pages} {22A511} (\bibinfo {year} {2012})}\BibitemShut {NoStop}%
\bibitem [{\citenamefont {Ma}\ \emph {et~al.}(2012)\citenamefont {Ma}, \citenamefont {Zhu}, \citenamefont {Guo},\ and\ \citenamefont {Yarkony}}]{Ma137:JCP12}%
  \BibitemOpen
  \bibfield  {author} {\bibinfo {author} {\bibfnamefont {J.}~\bibnamefont {Ma}}, \bibinfo {author} {\bibfnamefont {X.}~\bibnamefont {Zhu}}, \bibinfo {author} {\bibfnamefont {H.}~\bibnamefont {Guo}},\ and\ \bibinfo {author} {\bibfnamefont {D.~R.}\ \bibnamefont {Yarkony}},\ }\bibfield  {title} {\bibinfo {title} {First principles determination of the {NH$_2$}/{ND$_2$}{($\tilde{A}$,$\tilde{X}$)} branching ratios for photodissociation of {NH$_3$}/{ND$_3$} via full-dimensional quantum dynamics based on a new quasi-diabatic representation of coupled ab initio potential energy surfaces},\ }\href {https://doi.org/10.1063/1.4753425} {\bibfield  {journal} {\bibinfo  {journal} {J. Chem. Phys.}\ }\textbf {\bibinfo {volume} {137}},\ \bibinfo {pages} {22A541} (\bibinfo {year} {2012})}\BibitemShut {NoStop}%
\bibitem [{\citenamefont {Zotev}\ \emph {et~al.}(2020)\citenamefont {Zotev}, \citenamefont {Moreno~Carrascosa}, \citenamefont {Simmermacher},\ and\ \citenamefont {Kirrander}}]{Zotev16:JCTC20}%
  \BibitemOpen
  \bibfield  {author} {\bibinfo {author} {\bibfnamefont {N.}~\bibnamefont {Zotev}}, \bibinfo {author} {\bibfnamefont {A.}~\bibnamefont {Moreno~Carrascosa}}, \bibinfo {author} {\bibfnamefont {M.}~\bibnamefont {Simmermacher}},\ and\ \bibinfo {author} {\bibfnamefont {A.}~\bibnamefont {Kirrander}},\ }\bibfield  {title} {\bibinfo {title} {Excited electronic states in total isotropic scattering from molecules},\ }\href {https://doi.org/10.1021/acs.jctc.9b00670} {\bibfield  {journal} {\bibinfo  {journal} {J. Chem. Theory Comput.}\ }\textbf {\bibinfo {volume} {16}},\ \bibinfo {pages} {2594} (\bibinfo {year} {2020})}\BibitemShut {NoStop}%
\bibitem [{\citenamefont {Rodríguez}\ \emph {et~al.}(2014)\citenamefont {Rodríguez}, \citenamefont {González}, \citenamefont {Rubio-Lago},\ and\ \citenamefont {Bañares}}]{Rodríguez16:PCCP14}%
  \BibitemOpen
  \bibfield  {author} {\bibinfo {author} {\bibfnamefont {J.~D.}\ \bibnamefont {Rodríguez}}, \bibinfo {author} {\bibfnamefont {M.~G.}\ \bibnamefont {González}}, \bibinfo {author} {\bibfnamefont {L.}~\bibnamefont {Rubio-Lago}},\ and\ \bibinfo {author} {\bibfnamefont {L.}~\bibnamefont {Bañares}},\ }\bibfield  {title} {\bibinfo {title} {A velocity map imaging study of the photodissociation of the Ã state of ammonia},\ }\href {https://doi.org/10.1039/C3CP53523A} {\bibfield  {journal} {\bibinfo  {journal} {Phys. Chem. Chem. Phys.}\ }\textbf {\bibinfo {volume} {16}},\ \bibinfo {pages} {406} (\bibinfo {year} {2014})}\BibitemShut {NoStop}%
\bibitem [{\citenamefont {Xie}\ \emph {et~al.}(2015)\citenamefont {Xie}, \citenamefont {Zhu}, \citenamefont {Ma}, \citenamefont {Yarkony}, \citenamefont {Xie},\ and\ \citenamefont {Guo}}]{Xie142:JCP15}%
  \BibitemOpen
  \bibfield  {author} {\bibinfo {author} {\bibfnamefont {C.}~\bibnamefont {Xie}}, \bibinfo {author} {\bibfnamefont {X.}~\bibnamefont {Zhu}}, \bibinfo {author} {\bibfnamefont {J.}~\bibnamefont {Ma}}, \bibinfo {author} {\bibfnamefont {D.~R.}\ \bibnamefont {Yarkony}}, \bibinfo {author} {\bibfnamefont {D.}~\bibnamefont {Xie}},\ and\ \bibinfo {author} {\bibfnamefont {H.}~\bibnamefont {Guo}},\ }\bibfield  {title} {\bibinfo {title} {Communication: On the competition between adiabatic and nonadiabatic dynamics in vibrationally mediated ammonia photodissociation in its {A} band},\ }\href {https://doi.org/10.1063/1.4913633} {\bibfield  {journal} {\bibinfo  {journal} {J. Chem. Phys.}\ }\textbf {\bibinfo {volume} {142}},\ \bibinfo {pages} {091101} (\bibinfo {year} {2015})}\BibitemShut {NoStop}%
\bibitem [{\citenamefont {Hause}\ \emph {et~al.}(2006)\citenamefont {Hause}, \citenamefont {Yoon},\ and\ \citenamefont {Crim}}]{Hause12:JCP06}%
  \BibitemOpen
  \bibfield  {author} {\bibinfo {author} {\bibfnamefont {M.~L.}\ \bibnamefont {Hause}}, \bibinfo {author} {\bibfnamefont {Y.~H.}\ \bibnamefont {Yoon}},\ and\ \bibinfo {author} {\bibfnamefont {F.~F.}\ \bibnamefont {Crim}},\ }\bibfield  {title} {\bibinfo {title} {Vibrationally mediated photodissociation of ammonia: The influence of {N–H} stretching vibrations on passage through conical intersections},\ }\href {https://doi.org/10.1063/1.2363192} {\bibfield  {journal} {\bibinfo  {journal} {J. Chem. Phys.}\ }\textbf {\bibinfo {volume} {125}},\ \bibinfo {pages} {174309} (\bibinfo {year} {2006})}\BibitemShut {NoStop}%
\bibitem [{\citenamefont {McCarthy}\ \emph {et~al.}(1987)\citenamefont {McCarthy}, \citenamefont {Rosmus}, \citenamefont {Werner}, \citenamefont {Botschwina},\ and\ \citenamefont {Vaida}}]{McCarthy86:JCP87}%
  \BibitemOpen
  \bibfield  {author} {\bibinfo {author} {\bibfnamefont {M.~I.}\ \bibnamefont {McCarthy}}, \bibinfo {author} {\bibfnamefont {P.}~\bibnamefont {Rosmus}}, \bibinfo {author} {\bibfnamefont {H.}~\bibnamefont {Werner}}, \bibinfo {author} {\bibfnamefont {P.}~\bibnamefont {Botschwina}},\ and\ \bibinfo {author} {\bibfnamefont {V.}~\bibnamefont {Vaida}},\ }\bibfield  {title} {\bibinfo {title} {Dissociation of {NH$_3$} to {NH$_2$+H}},\ }\href {https://doi.org/10.1063/1.452417} {\bibfield  {journal} {\bibinfo  {journal} {J. Chem. Phys.}\ }\textbf {\bibinfo {volume} {86}},\ \bibinfo {pages} {6693} (\bibinfo {year} {1987})}\BibitemShut {NoStop}%
\bibitem [{\citenamefont {Li}\ and\ \citenamefont {Varandas}(2010)}]{Li114:JPCA10}%
  \BibitemOpen
  \bibfield  {author} {\bibinfo {author} {\bibfnamefont {Y.~Q.}\ \bibnamefont {Li}}\ and\ \bibinfo {author} {\bibfnamefont {A.~J.~C.}\ \bibnamefont {Varandas}},\ }\bibfield  {title} {\bibinfo {title} {{Ab-Initio}-based global double many-body expansion potential energy surface for the electronic ground state of the ammonia molecule},\ }\href {https://doi.org/10.1021/jp1019685} {\bibfield  {journal} {\bibinfo  {journal} {J. Phys. Chem. A}\ }\textbf {\bibinfo {volume} {114}},\ \bibinfo {pages} {6669} (\bibinfo {year} {2010})}\BibitemShut {NoStop}%
\bibitem [{\citenamefont {Yang}\ \emph {et~al.}(2021{\natexlab{a}})\citenamefont {Yang}, \citenamefont {Nunes}, \citenamefont {Ledbetter}, \citenamefont {Biasin}, \citenamefont {Centurion}, \citenamefont {Chen}, \citenamefont {Cordones}, \citenamefont {Crissman}, \citenamefont {Deponte}, \citenamefont {Glenzer}, \citenamefont {Lin}, \citenamefont {Mo}, \citenamefont {Rankine}, \citenamefont {Shen}, \citenamefont {Wolf},\ and\ \citenamefont {Wang}}]{YangJ23:PCCP21}%
  \BibitemOpen
  \bibfield  {author} {\bibinfo {author} {\bibfnamefont {J.}~\bibnamefont {Yang}}, \bibinfo {author} {\bibfnamefont {J.~P.~F.}\ \bibnamefont {Nunes}}, \bibinfo {author} {\bibfnamefont {K.}~\bibnamefont {Ledbetter}}, \bibinfo {author} {\bibfnamefont {E.}~\bibnamefont {Biasin}}, \bibinfo {author} {\bibfnamefont {M.}~\bibnamefont {Centurion}}, \bibinfo {author} {\bibfnamefont {Z.}~\bibnamefont {Chen}}, \bibinfo {author} {\bibfnamefont {A.~A.}\ \bibnamefont {Cordones}}, \bibinfo {author} {\bibfnamefont {C.}~\bibnamefont {Crissman}}, \bibinfo {author} {\bibfnamefont {D.~P.}\ \bibnamefont {Deponte}}, \bibinfo {author} {\bibfnamefont {S.~H.}\ \bibnamefont {Glenzer}}, \bibinfo {author} {\bibfnamefont {M.-F.}\ \bibnamefont {Lin}}, \bibinfo {author} {\bibfnamefont {M.}~\bibnamefont {Mo}}, \bibinfo {author} {\bibfnamefont {C.~D.}\ \bibnamefont {Rankine}}, \bibinfo {author} {\bibfnamefont {X.}~\bibnamefont {Shen}}, \bibinfo {author} {\bibfnamefont {T.~J.~A.}\ \bibnamefont {Wolf}},\ and\ \bibinfo {author} {\bibfnamefont
  {X.}~\bibnamefont {Wang}},\ }\bibfield  {title} {\bibinfo {title} {Structure retrieval in liquid-phase electron scattering},\ }\href@noop {} {\bibfield  {journal} {\bibinfo  {journal} {Phys. Chem. Chem. Phys.}\ }\textbf {\bibinfo {volume} {23}},\ \bibinfo {pages} {1308} (\bibinfo {year} {2021}{\natexlab{a}})}\BibitemShut {NoStop}%
\bibitem [{\citenamefont {Cheng}\ \emph {et~al.}(2006)\citenamefont {Cheng}, \citenamefont {Lu}, \citenamefont {Chen}, \citenamefont {Bahou}, \citenamefont {Lee}, \citenamefont {Mebel}, \citenamefont {Lee}, \citenamefont {Liang},\ and\ \citenamefont {Yung}}]{Cheng647:Astro06}%
  \BibitemOpen
  \bibfield  {author} {\bibinfo {author} {\bibfnamefont {B.-M.}\ \bibnamefont {Cheng}}, \bibinfo {author} {\bibfnamefont {H.-C.}\ \bibnamefont {Lu}}, \bibinfo {author} {\bibfnamefont {H.-K.}\ \bibnamefont {Chen}}, \bibinfo {author} {\bibfnamefont {M.}~\bibnamefont {Bahou}}, \bibinfo {author} {\bibfnamefont {Y.-P.}\ \bibnamefont {Lee}}, \bibinfo {author} {\bibfnamefont {A.~M.}\ \bibnamefont {Mebel}}, \bibinfo {author} {\bibfnamefont {L.~C.}\ \bibnamefont {Lee}}, \bibinfo {author} {\bibfnamefont {M.-C.}\ \bibnamefont {Liang}},\ and\ \bibinfo {author} {\bibfnamefont {Y.~L.}\ \bibnamefont {Yung}},\ }\bibfield  {title} {\bibinfo {title} {Absorption cross sections of {NH$_3$}, {NH$_2$D}, {NHD$_2$}, and {ND$_3$} in the spectral range 140-220 nm and implications for planetary isotopic fractionation},\ }\href {https://doi.org/10.1086/505615} {\bibfield  {journal} {\bibinfo  {journal} {Astrophys. J.}\ }\textbf {\bibinfo {volume} {647}},\ \bibinfo {pages} {1535} (\bibinfo {year} {2006})}\BibitemShut {NoStop}%
\bibitem [{\citenamefont {Yang}\ \emph {et~al.}(2021{\natexlab{b}})\citenamefont {Yang}, \citenamefont {Dettori}, \citenamefont {Nunes}, \citenamefont {List}, \citenamefont {Biasin}, \citenamefont {Centurion}, \citenamefont {Chen}, \citenamefont {Cordones}, \citenamefont {Deponte}, \citenamefont {Heinz} \emph {et~al.}}]{YangJ21:Nature596}%
  \BibitemOpen
  \bibfield  {author} {\bibinfo {author} {\bibfnamefont {J.}~\bibnamefont {Yang}}, \bibinfo {author} {\bibfnamefont {R.}~\bibnamefont {Dettori}}, \bibinfo {author} {\bibfnamefont {J.~P.~F.}\ \bibnamefont {Nunes}}, \bibinfo {author} {\bibfnamefont {N.~H.}\ \bibnamefont {List}}, \bibinfo {author} {\bibfnamefont {E.}~\bibnamefont {Biasin}}, \bibinfo {author} {\bibfnamefont {M.}~\bibnamefont {Centurion}}, \bibinfo {author} {\bibfnamefont {Z.}~\bibnamefont {Chen}}, \bibinfo {author} {\bibfnamefont {A.~A.}\ \bibnamefont {Cordones}}, \bibinfo {author} {\bibfnamefont {D.~P.}\ \bibnamefont {Deponte}}, \bibinfo {author} {\bibfnamefont {T.~F.}\ \bibnamefont {Heinz}}, \emph {et~al.},\ }\bibfield  {title} {\bibinfo {title} {Direct observation of ultrafast hydrogen bond strengthening in liquid water},\ }\href@noop {} {\bibfield  {journal} {\bibinfo  {journal} {Nature}\ }\textbf {\bibinfo {volume} {596}},\ \bibinfo {pages} {531} (\bibinfo {year} {2021}{\natexlab{b}})}\BibitemShut {NoStop}%
\bibitem [{\citenamefont {Centurion}\ \emph {et~al.}(2022)\citenamefont {Centurion}, \citenamefont {Wolf},\ and\ \citenamefont {Yang}}]{PDF}%
  \BibitemOpen
  \bibfield  {author} {\bibinfo {author} {\bibfnamefont {M.}~\bibnamefont {Centurion}}, \bibinfo {author} {\bibfnamefont {T.~J.}\ \bibnamefont {Wolf}},\ and\ \bibinfo {author} {\bibfnamefont {J.}~\bibnamefont {Yang}},\ }\bibfield  {title} {\bibinfo {title} {Ultrafast imaging of molecules with electron diffraction},\ }\href {https://doi.org/https://doi.org/10.1146/annurev-physchem-082720-010539} {\bibfield  {journal} {\bibinfo  {journal} {Annu. Rev. Phys. Chem.}\ }\textbf {\bibinfo {volume} {73}},\ \bibinfo {pages} {21} (\bibinfo {year} {2022})}\BibitemShut {NoStop}%
\bibitem [{\citenamefont {Wells}\ \emph {et~al.}(2009)\citenamefont {Wells}, \citenamefont {Perriam},\ and\ \citenamefont {Stavros}}]{Wells130:JCP09}%
  \BibitemOpen
  \bibfield  {author} {\bibinfo {author} {\bibfnamefont {K.~L.}\ \bibnamefont {Wells}}, \bibinfo {author} {\bibfnamefont {G.}~\bibnamefont {Perriam}},\ and\ \bibinfo {author} {\bibfnamefont {V.~G.}\ \bibnamefont {Stavros}},\ }\bibfield  {title} {\bibinfo {title} {Time-resolved velocity map ion imaging study of {NH$_3$} photodissociation},\ }\href {https://doi.org/10.1063/1.3072763} {\bibfield  {journal} {\bibinfo  {journal} {J. Chem. Phys.}\ }\textbf {\bibinfo {volume} {130}},\ \bibinfo {pages} {074308} (\bibinfo {year} {2009})}\BibitemShut {NoStop}%
\bibitem [{\citenamefont {Richter}\ \emph {et~al.}(2011)\citenamefont {Richter}, \citenamefont {Marquetand}, \citenamefont {González-Vázquez}, \citenamefont {Sola},\ and\ \citenamefont {González}}]{SHARC_Richter7:JCTC1253}%
  \BibitemOpen
  \bibfield  {author} {\bibinfo {author} {\bibfnamefont {M.}~\bibnamefont {Richter}}, \bibinfo {author} {\bibfnamefont {P.}~\bibnamefont {Marquetand}}, \bibinfo {author} {\bibfnamefont {J.}~\bibnamefont {González-Vázquez}}, \bibinfo {author} {\bibfnamefont {I.}~\bibnamefont {Sola}},\ and\ \bibinfo {author} {\bibfnamefont {L.}~\bibnamefont {González}},\ }\bibfield  {title} {\bibinfo {title} {{SHARC}: ab initio molecular dynamics with surface hopping in the adiabatic representation including arbitrary couplings},\ }\href@noop {} {\bibfield  {journal} {\bibinfo  {journal} {J. Chem. Theory Comput.}\ }\textbf {\bibinfo {volume} {7}},\ \bibinfo {pages} {1253} (\bibinfo {year} {2011})}\BibitemShut {NoStop}%
\bibitem [{\citenamefont {Seritan}\ \emph {et~al.}(2020)\citenamefont {Seritan}, \citenamefont {Bannwarth}, \citenamefont {Fales}, \citenamefont {Hohenstein}, \citenamefont {Kokkila-Schumacher}, \citenamefont {Luehr}, \citenamefont {Snyder}, \citenamefont {Song}, \citenamefont {Titov}, \citenamefont {Ufimtsev},\ and\ \citenamefont {Martínez}}]{TERACHEM_Seritan152:JCTC20}%
  \BibitemOpen
  \bibfield  {author} {\bibinfo {author} {\bibfnamefont {S.}~\bibnamefont {Seritan}}, \bibinfo {author} {\bibfnamefont {C.}~\bibnamefont {Bannwarth}}, \bibinfo {author} {\bibfnamefont {B.~S.}\ \bibnamefont {Fales}}, \bibinfo {author} {\bibfnamefont {E.~G.}\ \bibnamefont {Hohenstein}}, \bibinfo {author} {\bibfnamefont {S.~I.~L.}\ \bibnamefont {Kokkila-Schumacher}}, \bibinfo {author} {\bibfnamefont {N.}~\bibnamefont {Luehr}}, \bibinfo {author} {\bibfnamefont {J.}~\bibnamefont {Snyder}, \bibfnamefont {James~W.}}, \bibinfo {author} {\bibfnamefont {C.}~\bibnamefont {Song}}, \bibinfo {author} {\bibfnamefont {A.~V.}\ \bibnamefont {Titov}}, \bibinfo {author} {\bibfnamefont {I.~S.}\ \bibnamefont {Ufimtsev}},\ and\ \bibinfo {author} {\bibfnamefont {T.~J.}\ \bibnamefont {Martínez}},\ }\bibfield  {title} {\bibinfo {title} {{TeraChem}: Accelerating electronic structure and ab initio molecular dynamics with graphical processing units},\ }\href@noop {} {\bibfield  {journal} {\bibinfo  {journal} {J. Chem. Phys.}\ }\textbf
  {\bibinfo {volume} {152}},\ \bibinfo {pages} {224110} (\bibinfo {year} {2020})}\BibitemShut {NoStop}%
\end{thebibliography}%

\clearpage
\begin{figure*}[h!]
    \centering
    \includegraphics[width=0.9\linewidth]{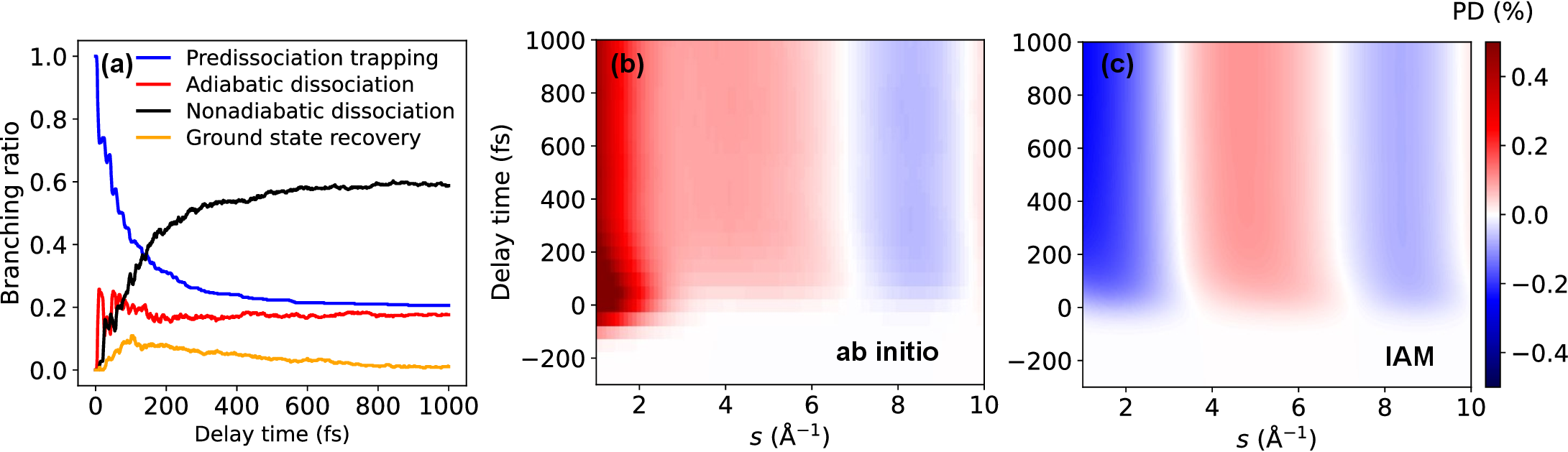}
    \caption{(a) Time-dependent populations of different reaction channels obtained from MD simulations. (b) Time-resolved \textit{ab initio} PD obtained from MD simulations. (c) Time-resolved PD calculated using the IAM model along trajectories obtained from MD simulations. 
    }
    \label{fig:PD_IAM_ab}
\end{figure*}

\textbf{End Matter on Calculation of CPDF from diffraction signals.}
CPDF is defined as~\cite{YangJ23:PCCP21}
\begin{equation}
    \text{CPDF}(r) = \sum_{uv} Z_{u} Z_{v} P_{uv}(r)\,,
\end{equation}
where the sum runs over all charge-pairs, including electron-nucleus, nucleus-nucleus, and electron-electron pairs as shown in Fig.~\ref{fig:schematic}(b), \( Z_u \) and \( Z_v \) are their charges, and \( P_{uv}(r) \) represents the probability distribution of finding the particle pairs \( u \) and \( v \) separated by distance \( r \). The electron diffraction intensity is
\begin{equation}
    I(s) = \frac{1}{s^{4}} \left[ \sum_{uv}  Z_{u} Z_{v} \int P_{uv}(r) \frac{\sin(sr)}{sr} \, dr \right]\,,
    \label{totalI}
\end{equation}
where $s$ is the momentum transfer of the incident electron. CPDF is obtained by taking the Fourier sine transform of $I(s)$
\begin{equation}
    \text{CPDF}(r) = r \int_{s_{\text{min}}}^{s_{\text{max}}} s^{5} I(s) e^{-\alpha s^{2}} \sin(sr) \, ds\,,
    \label{CPDF_INVERSION}
\end{equation}
where the damping term $e^{-\alpha s^{2}}$ is added to eliminate edge effects due to the finite $s$ range.
In this work, both theoretical and experimental $\Delta$CPDF is computed using the same $s$-range 1--10~$\text{\AA}$$^{-1}$ and the damping parameter $\alpha=0.08$~\AA$^{2}$.

\textbf{End Matter on Molecular dynamics simulations and \textit{ab initio} calculation of diffraction signals.}
We simulate photodissociation dynamics of NH$_3$ molecules after pump pulse excitation by the fewest-switches surface hopping method using SHARC package~\cite{SHARC_Richter7:JCTC1253}. The MD simulation is carried out based on the two-state Hamiltonian model established in Ref.~\cite{Zhu137:JCP12}, which provides the adiabatic potential energy surface and derivative couplings. All MD trajectories are classified to the four reaction channels shown in Fig.~\ref{fig:schematic}(a) based on their N--H bond length and electronic state~\cite{ThomasWolf131:PRL23}, and the temporal evolution of the branching ratios are shown in Fig.~\ref{fig:PD_IAM_ab}(a). Initially all NH$_3$ molecules are excited to the $S_1$ state followed by N--H bond elongation and umbrella vibration. At 1 ps, $\sim$60\% of the NH$_3$ molecules undergoes a transition to the ground state via the conical intersection and dissociate to NH$_2$+H. For the remaining molecules in the excited state $S_1$, $\sim$20\% dissociates adiabatically, and the other $\sim$20\% are confined behind the potential barrier before dissociation in $S_1$.

The \textit{ab initio} elastic and inelastic scattering signals are calculated based on the MD trajectories using TeraChem package~\cite{TERACHEM_Seritan152:JCTC20}. The electronic state is calculated based on the complete active space self-consistent field (CASSCF) theory. The active space consists of 8 electrons and 8 orbitals, and the basis set aug-cc-pVDZ is used. The rotationally averaged scattering signals are calculated using a 590-point Lebedev quadrature. The \textit{ab initio} calculated PD signals are shown in Fig.~\ref{fig:PD_IAM_ab}(b).
Fig.~\ref{fig:PD_IAM_ab}(c) shows the diffraction intensity calculated within the IAM model
\begin{eqnarray}
    I_{\mathrm{IAM}}(s)=\sum_I f^2_I(s)+\sum_{I\neq J} f_I(s)f_J(s)\frac{\sin(s r_{IJ})}{s r_{IJ}}\,,
\end{eqnarray}
where $I,J$ are atomic indexes in NH$_3$ molecule, $f_I(s)$ is the form factor of the $I$-th atom, and $r_{IJ}$ is the distance between the $I$-th atom and $J$-th atom.

\textbf{End Matter on Extracting the experimental inelastic scattering signals.}
\begin{figure}
    \centering
    \includegraphics[width=0.9\linewidth]{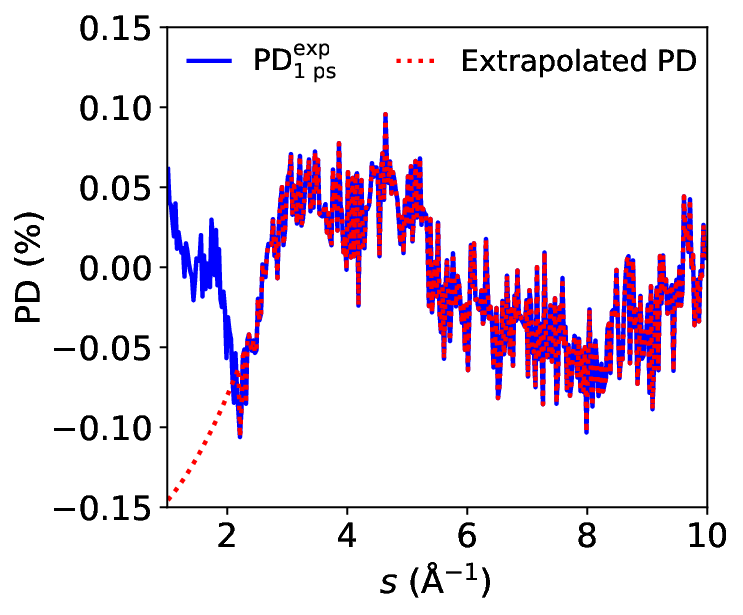}
    \caption{The experimental PD at 1~ps, before (blue) and after (red) extrapolation. The extrapolation is performed using a low-order polynomial fit to the experimental data within 2.2--3.3~$\text{\AA}^{-1}$.
    }
    \label{fig:extrapolation}
\end{figure}
In Fig.~\ref{fig:extrapolation}, we present the experimentally measured PD around 1~ps, in which pronounced negative signals appear near 2~$\text{\AA}^{-1}$. However, this signal is strongly affected by the inelastic contribution at $s<2~\text{\AA}^{-1}$, making it challenging to separate the elastic and inelastic signals in the small-angle region. Also, we notice that the branching ratios obtained from \textit{ab initio} calculations do not fully agree with the experimental results, as discussed in the main text. This discrepancy affects the theoretical signal in the small scattering angle region--a known issue also reported in previous UED studies~\cite{ThomasWolf131:PRL23}, where theoretical predictions at the low-$s$ region deviate from experimental observations. Therefore, we choose to use the experimental data to separate the elastic and inelastic contributions in the small-angle region, instead of using the simulation results.
%We find that the experimental PD near 2~$\text{\AA}^{-1}$ exhibits a steep rise as $s$ decreases, indicating a strong influence from the inelastic signal in this region. 
As the inelastic contribution diminishes rapidly with increasing $s$, the elastic signals that extend beyond 2~$\text{\AA}^{-1}$ are considerably less impacted. Therefore, we extrapolate the signal in the small-angle region by fitting a smooth low-order polynomial to the measured PD values in the range of 2.2--3.3~$\text{\AA}^{-1}$. The extrapolated result is shown in Fig.~\ref{fig:extrapolation}. We find that the extrapolated elastic signal at 1~$\text{\AA}^{-1}$ closely matches the IAM results shown in Fig.~\ref{fig:PD_IAM_ab}(c) ($-$0.15\%), further supporting the reliability of our extrapolation. It should be noted that this extrapolation is solely used to assist in fitting the decay constant of the inelastic signal in Fig.~\ref{fig:PD_exp}(b). In the CPDF calculations, we use the complete original experimental data, which includes both elastic and inelastic contributions.

%\section{Third Appendix Section}

%\begin{figure}
%    \centering
%    \includegraphics[width=0.9\linewidth]{F_deltadeltaCPDF.eps}
%    \caption{Difference in $\Delta$CPDF between the nonadiabatic and adiabatic dissociation Channels %($\Delta$CPDF$_{\text{nonadi}}$-$\Delta$CPDF$_{\text{adi}}$). The red, black, blue, and orange lines represent the total difference, and the ee, eN, and NN components, respectively.
%    }
%    \label{fig:deltadeltaCPDF}
%\end{figure}

\end{document}